\documentclass[10pt]{iopart}
\usepackage{epsfig,graphicx}
\usepackage{amsfonts}

\newcommand{\sigmav}{\mbox{\boldmath$\sigma$}}

\newcommand{\hv}{\mbox{\boldmath$h$}}
\newcommand{\Hv}{\mbox{\boldmath$H$}}
\newcommand{\xv}{\mbox{\boldmath$x$}}
\newcommand{\thetav}{\mbox{\boldmath$\theta$}}

\newcommand{\bra}{\langle}
\newcommand{\ket}{\rangle}

\newcommand{\minus}{{\!-\!}}

\newcommand{\be}{\begin{equation}}
\newcommand{\ee}{\end{equation}}
\newcommand{\bd}{\begin{displaymath}}
\newcommand{\ed}{\end{displaymath}}

\newcommand{\room}{\rule[-0.1cm]{0cm}{0.6cm}}
\newcommand{\bigroom}{\rule[-0.4cm]{0cm}{1.0cm}}

\newcommand{\Z}{{\mathcal Z}}

\newcommand{\bx}{\ensuremath{\mathbf{x}}}
\newcommand{\by}{\ensuremath{\mathbf{y}}}

\newcommand{\here}{\makebox(0,0)}

\newcommand{\cv}{\mbox{\boldmath$c$}}

\begin{document}

\title[Dynamics in the Griffiths phase of the diluted Ising ferromagnet]{Dynamical replica analysis of processes on finitely connected random graphs II: Dynamics in the Griffiths phase of the diluted Ising ferromagnet}
\author{A Mozeika and ACC Coolen}
\address{
Department of Mathematics, King's College London\\ The Strand,
London WC2R 2LS, UK }


\ead{alexander.mozeika@kcl.ac.uk,ton.coolen@kcl.ac.uk}



\begin{abstract}
We study the Glauber dynamics of Ising spin models with random bonds, on finitely connected random graphs.
We generalize a recent dynamical replica theory with which to predict the evolution of the joint spin-field distribution, to include random graphs with arbitrary degree distributions. The theory is applied to Ising ferromagnets on randomly diluted Bethe lattices, where  we study the evolution of the magnetization and the internal energy. It predicts a prominent slowing down of the flow in the Griffiths phase, it suggests a further dynamical transition at lower temperatures within the Griffiths phase, and it is verified quantitatively by the results of Monte Carlo simulations.
\end{abstract}



\section{Introduction}\label{section2:Intro}
Finitely connected (FC) spin systems were introduced more than 20 years ago by Viana and Bray \cite{Viana} as more realistic alternatives to infinite range (IR) models of spin-glasses \cite{SK,BOOK}. In finitely connected systems the spins are placed on the vertices of a random graph,  and interact only when their vertices are connected; the number of connections per spin remains finite (on average), even in the thermodynamic limit.
This definition endows finitely connected spin models with a geometry (e.g. vertex neighborhood), a
crucial feature also of finite dimensional (FD) spin systems, that was absent from infinite range models.
Yet, in contrast to FD spin systems which are notoriously difficult to solve, FC models are still of a mean field nature and can therefore be studied analytically using methods from the statistical mechanics of disordered systems. This property reflects the absence of short loops: in finitely connected spin systems loop lengths are typically of order $\log(N)$,  so that the spins live in environments which are locally tree-like, unlike spins in finite dimensional systems, and short-range frustration cannot occur. As a result of their analytical accessability the equilibrium properties of finitely connected spin systems are now understood quite well \cite{Kanter,BetheSG1,Monasson,BetheSG,SoftSpins,PerezVicenteCoolen}. The mathematical and numerical techniques which originated from these equilibrium papers were, in turn, generalized and applied in subsequent dynamical studies \cite{diluteDyn,StochDyn,ApprSch,ParDyn,DRTfc,Goos,LLmodel,MozeikaCoolen,KiemesHorner}.

One of the properties shared by finitely connected and finite dimensional spin systems is the presence of Griffiths singularities \cite{Griffiths}. In his seminal paper Griffiths showed that in the diluted Ising ferromagnet, where either sites or bonds of a classical lattice are removed with some probability $1-p$, the magnetization is a non-analytical function of the external field for a range of temperatures $T_c(p)<T<T_c(1)$, where $T_c(p)$ and  $T_c(1)$ are the critical temperatures marking the P$\to$F transition of the diluted and undiluted systems, respectively. The system is in a conventional paramagnetic state only for temperatures above $T_c(1)$, where $T_c(1)$ could be infinite \cite{BrayHuifang}. The temperature interval  $T_c(p)<T<T_c(1)$ over which these singularities occur is called the Griffiths phase \cite{Randeria}.
This peculiar behavior of the magnetization \cite{Harris} and other thermodynamic functions is understood to be caused
 by the presence in the randomly diluted system of large undiluted spatial regions (or clusters) of the lattice. In the Griffiths phase these clusters are in an ordered magnetic state, although the system is globally paramagnetic. The Griffiths singularities are not always strong\footnote{See \cite{Hinczewski} for a model example where very strong Griffiths effects are found.} \cite{Harris,Imry,Schwartz,BrayHuifang} and often difficult to observe experimentally \cite{Imry}, nevertheless this is possible with modern sampling techniques \cite{HukushimaIba}.

In contrast to statics, the effects of large undiluted clusters on the dynamic properties of diluted spin systems are more drastic. The dynamics in such clusters is very slow because it requires reversing spins coherently in the entire cluster. In FD spin systems this results in non-exponential decay of the spin autocorrelation and magnetization functions in the entire Griffiths region \cite{Randeria,Nature,BrayDynamics,Bounds,ColborneBray,Jain}. The latter studies concentrated mainly on the derivation of bounds for the spin autocorrelation function, at large times, with subsequent verification by Monte Carlo (MC) simulations. The dynamic properties of the Griffiths phase in FC spin systems remain, to the best of our knowledge and that of others \cite{Vojta}, largely unexplored.

In this paper we generalize recent results obtained within the framework of dynamic replica theory (DRT) \cite{MozeikaCoolen} to include random graphs with arbitrary vertex degree distributions, and apply the generalized theory to the dynamics of diluted ferromagnets in the Griffiths phase. Our paper is organized as follows. In section \ref{section2:Model} we define our finitely connected spin model and its dynamical equations. In section \ref{section2:RA} we close the macroscopic dynamical laws using the standard assumptions and procedures of DRT. From these closed laws we recover known results of equilibrium statistical mechanics as stationary solutions in section \ref{section2:Equilibr}, as a test. The replica-symmetry assumption allows us to take the replica limit $n\to 0$ in section \ref{section2:RS}. The resulting dynamical theory is applied to the Glauber dynamics of diluted Ising ferromagnet in section \ref{section2:Griffiths}. We close with a summary and discussion of our results.

\section{Model definitions and dynamic equations}\label{section2:Model}
\indent We  consider a system of $N$ Ising spins $\sigma_i
\in\{-1,1\}$, which are placed on the vertices of a finitely connected random graph. Spins interact only when they are connected. Their microscopic dynamics is described by a master
equation for the evolution of the microscopic state probability
in continuous time:
\begin{eqnarray}
\frac{\mathrm{d}}{\mathrm{d}t}p_t(\sigmav) &=& \sum_{i=1}^N [p_t(F_i
\sigmav)w_i(F_i \sigmav)-
p_t(\sigmav)w_i(\sigmav)]\label{eq2:master}
\end{eqnarray}
in which $\sigmav=(\sigma_1,\ldots,\sigma_N)$, $F_i$ denotes the spin-flip operator defined via $F_i \Omega (\sigmav)=\Omega
(\sigma_1,\ldots,-\sigma_i,\ldots,\sigma_N)$, and the quantities
$w_i(\sigmav)$ are the Glauber transition rates
\begin{eqnarray}
w_i(\sigmav)&=&\frac{1}{2}[1-\sigma_i \tanh[\beta h_i(\sigmav)]]\label{eq2:rate}
\end{eqnarray}
with the local fields
 \begin{eqnarray}h_i(\sigmav)=\sum_{j\neq i} c_{ij}J_{ij}\sigma_j+
\theta\label{eq2:field}.
\end{eqnarray}
The parameters $\beta=T^{-1}$ and $\theta$ define the inverse temperature and a uniform external field, respectively. The random interactions $\{c_{ij}J_{ij}\}$ are symmetric,
viz. $c_{ij}J_{ij}=c_{ji}J_{ji}$, and are regarded as a quenched disorder. The interaction strengths $J_{ij}$ are independent random variables, drawn from a probability distribution $P(J)$. The random variables $c_{ij}\in \{0,1\}$ are the entries of an adjacency matrix, with zeroes on the main diagonal, defining the random interaction graph. The symmetry of the interactions ensures that the process (\ref{eq2:master}) evolves towards equilibrium, characterized by the Boltzmann measure $p_\infty (\sigmav)\sim \exp[-\beta H(\sigmav)]$,
with the Hamiltonian
\begin{eqnarray}
H(\sigmav)=-\sum_{i<j}\sigma_i
c_{ij}J_{ij}\sigma_j-\theta\sum_i \sigma_i~.\label{def2:E}
\end{eqnarray}
In this paper we consider FC random
graphs where the vertex degrees $\lbrace k_i\rbrace$, with $k_i=\sum_{j\neq i}c_{ij}$, are drawn randomly and independently from an arbitrary probability distribution $P_c(k)$ over the non-negative integers, with finite first moment $c=\sum_k P_c(k)k$. The probability of finding an adjacency matrix $\cv=\{c_{ij}\}$ in this random graph ensemble, constrained by the vertex degrees $\lbrace k_i\rbrace$, is given by
\begin{eqnarray}
P(\cv\vert\{k_i\})&=&\frac{1}{\Z}\prod_{i<j}p_c(c_{ij})\prod_i\delta_{k_i,\sum_{j\neq i} c_{ij}}\label{eq2:P(C)}
\end{eqnarray}
where $\Z$ is a normalization constant, and
\begin{eqnarray}
\forall i<j:&~~~& p_c(c_{ij})=\frac{c}{N}\delta_{c_{ij},1}+
(1-\frac{c}{N})\delta_{c_{ij},0}~.\label{eq2:P(c)}
\end{eqnarray}
The presence of $p_c(c_{ij})$ in the definition (\ref{eq2:P(C)}) is mathematically convenient in solving the model, but not essential; it can be transformed away in leading order in $N$.

We avoid the impossible task of solving the $2^N$ equations (\ref{eq2:master}) directly, and consider an alternative description of the dynamics in terms of macroscopic observables. In particular, we consider the evolution in time of the joint spin-field distribution \cite{DRTinf2}, which is given by
\begin{eqnarray}
D(s,h;\sigmav)=\frac{1}{N}\sum_i\delta_{s,\sigma_i}\delta\left[h-
h_i(\sigmav)\right]. \label{def2:jsfield}
\end{eqnarray}
In finitely connected models equipped with the  dynamics (\ref{eq2:master}), the macroscopic distribution (\ref{def2:jsfield}) will evolve deterministically for $N\rightarrow\infty$, according to a macroscopic dynamical equation \cite{MozeikaCoolen}
of the form%
\begin{eqnarray}
\frac{\partial}{\partial t}D(s,h)&=& \frac{1}{2}\left [1+
s\tanh[\beta h]\right ]D(- s,h)-\frac{1}{2}\left [1-
s\tanh[\beta h]\right
]D(s,h) \nonumber
\\
&&\hspace{-1mm}+\frac{1}{2}c\sum_{{s^\prime}}\int\mathrm{d}{h^\prime}
[1-{s^\prime}\tanh[\beta {h^\prime}]]A[s,
s^\prime;h,
h^\prime;s^\prime]\nonumber\\
&&-\frac{1}{2}c\sum_{{s^\prime}}\int\mathrm{d}{h^\prime}
[1-{s^\prime}\tanh[\beta {h^\prime}]]A[s,
s^\prime;h, h^\prime;0]
\label{eq2:diffusion}
\end{eqnarray}
with a spin variable $s\in\{-1,1\}$,  and a field $h\in\mathbb{R}$.
  The dynamical equation (\ref{eq2:diffusion}) is written in terms of {\em time-dependent}  kernels $D$ and $A$, which are defined as follows
\begin{eqnarray}
D(s,h)&=&\frac{1}{N}\sum_i\langle\delta_{s,\sigma_i}\delta\left[h-
h_i(\sigmav)\right]\rangle_{D;t}\label{def2:D}\\
A[s,s^\prime;h,h^\prime;\tilde s]&=&\frac{1}{c
  N}\sum_{\ell,\ell^\prime}c_{\ell\ell^\prime}A_{\ell\ell^\prime}[s,s^\prime;h,h^\prime;\tilde s] \label{def2:A}\\
A_{\ell\ell^\prime}[s,s^\prime;h,h^\prime;\tilde s]&=&\Big\langle\delta_{{s^\prime},\sigma_\ell}\delta_{s,\sigma_{\ell^\prime}}\delta [
h^\prime\! -\! h_\ell (\sigmav)]\delta [ h\! -\! h_{\ell^\prime} (\sigmav)\!+\!
2J_{\ell\ell^\prime}\tilde s]\Big\rangle_{D;t}\nonumber
\end{eqnarray}
with $s^\prime\in\{-1,1\}$, $h^\prime\in\mathbb{R}$ and $\tilde s \in \{0,s^\prime\}$. In these expressions, the sub-shell average
\begin{eqnarray}
\left\langle f(\sigmav)\right\rangle_{D ; t}&=&
\frac{\sum_{\sigmav}p_t(\sigmav)f(\sigmav)\prod_{s h}\delta
\left[D(s,h) -
D(s,h;\sigmav)\right]}{\sum_{{\sigmav^\prime}}p_t({\sigmav^\prime})\prod_{s h}\delta
\left[D(s,h) -
D(s,h;{\sigmav^\prime})\right]}\label{def2:SSA}
\end{eqnarray}
is written in terms of the macroscopic distribution (\ref{def2:jsfield}) which acts as a constraint on micro-states\footnote{Here, to simplify notation, we skip explicit mentioning of the intermediate discretization of the fields $h$ in (\ref{def2:jsfield}), which is formally required \cite{MozeikaCoolen}.}, and the microscopic probability distribution $p_t(\sigmav)$. The kernel $A$ is positive semi-definite, and normalized for $N\rightarrow\infty$; it defines the joint spin-field probability distribution of connected sites. Equation (\ref{eq2:diffusion}) is exact for large $N$, but not yet closed due to the presence of microscopic probability $p_t(\sigmav)$ in (\ref{def2:SSA}).

\section{Dynamical replica analysis}\label{section2:RA}

\subsection{Closure and disorder averaging}

In order to solve equation (\ref{eq2:diffusion}) we have to compute the kernel (\ref{def2:A}). This latter kernel is dependent on the disorder $\left\{c_{ij}J_{ij}\right\}$ and the microscopic state probability $p_t(\sigmav)$. To compute $A$ we make the usual assumptions of the
 dynamic replica method \cite{DRTinf2,MozeikaCoolen}: (i) the observables $\{D(s,h;\sigmav)\}$ are
assumed to be self-averaging with respect to the disorder at any time, and  (ii) the
microscopic probability $p_t(\sigmav)$ is taken to depend on $\sigmav$ only through $\{D(s,h;\sigmav)\}$. The self-averaging assumption leads us to
\begin{eqnarray}
A[s,s^\prime;h,h^\prime;\tilde s]&=&\Big\langle\frac{1}{c
N}\sum_{\ell,\ell^\prime}c_{\ell\ell^\prime}A_{\ell\ell^\prime}[s,s^\prime;h,h^\prime;\tilde s]\Big\rangle_{\!\{c_{ij}J_{ij}\}}.\label{eq2:avA}
\end{eqnarray}
 The subsequent elimination of the microscopic probability $p_t(\sigmav)$ from the above, followed by the elimination of the fraction via the replica identity $\sum_{\sigmav}\Phi
(\sigmav)W(\sigmav)/\sum_{\sigmav^\prime}W(\sigmav^\prime)=\lim_{n\rightarrow
0}\sum_{\sigmav^1}\ldots\sum_{\sigmav^n}\Phi
(\sigmav^1)\prod_{\alpha=1 }^n W(\sigmav^{\alpha})$, allow us to perform the disorder averages  in the term $c_{\ell\ell^\prime}A_{\ell\ell^\prime}$ of equation (\ref{def2:A}) (see \ref{section2:average} for details), yielding
\begin{eqnarray}
\hspace*{-15mm}
&&\hspace{-15mm}\left\langle c_{\ell\ell^\prime}A_{\ell\ell^\prime}[s,s^\prime;h,h^\prime;\tilde s]\right\rangle_{\{c_{ij}J_{ij}\}}
\nonumber \\
\hspace*{-15mm}
&=&\sum_{\sigmav^1}\ldots\sum_{\sigmav^n}\int\prod_{\alpha
 i}\Big[ \mathrm d H_i^{\alpha}\mathrm d
\hat{h}_i^{\alpha}~
\exp[\rmi\hat{h}_i^{\alpha}H_i^{\alpha}]\Big]
\nonumber\\
\hspace*{-15mm}
&&\times\prod_{\tau h \alpha}\delta \Big[D(\tau,h)
-
\frac{1}{N}\sum_i\delta_{\tau,\sigma_i^{\alpha}}\delta[h-
H_i^{\alpha}]\Big]
\nonumber\\
\hspace*{-15mm}
&& \times\delta_{{s^\prime},\sigma_\ell^1}\delta_{s,\sigma_{\ell^\prime}^1}\delta
[h^\prime -H_\ell^1] ~\Big\langle c_{\ell\ell^\prime}\delta [ h -H_{\ell^\prime}^1+2J_{\ell\ell^\prime}\tilde s]\rme^{- \rmi\sum_{\alpha
i}\hat{h}_i^{\alpha}h_i
(\sigmav^{\alpha})}
\Big\rangle_{\{c_{ij}J_{ij}\}}
\nonumber\\
\hspace*{-15mm}
&=&\frac{1}{\Z}\frac{c}{N}\sum_{\sigmav^1}\ldots\sum_{\sigmav^n}\int\prod_{\alpha
, i}\left\{\frac{ \mathrm d H_i^{\alpha}\mathrm d
\hat{h}_i^{\alpha}}{2\pi}\right\}
\rme^{\rmi\sum_{\alpha , i}\hat{h}_i^{\alpha}H_i^{\alpha}-\rmi\theta\sum_{\alpha , i}\hat{h}_i^{\alpha}}\nonumber\\
\hspace*{-15mm}
&&
\times\prod_{\tau h \alpha}\delta
\left[D(\tau,h
) -\frac{1}{N}\sum_i\delta_{\tau,\sigma_i^{\alpha}}\delta\left[h
-H_i^{\alpha}\right]\right]
\nonumber\\
\hspace*{-15mm}
&&\times\int \mathrm d J~ P(J)~\delta_{s^\prime ,\sigma_\ell^1}\delta_{s,\sigma_{\ell^\prime}^1}\delta [ h
-H_{\ell^\prime}^1+2J\tilde s]\delta [h^\prime  -H_\ell^1]~\rme^{-\rmi J \sum_{\alpha}\big\{\hat{h}_\ell^{\alpha}
\sigma_{\ell^\prime}^{\alpha}+\hat{h}_{\ell^\prime}^{\alpha}
\sigma_\ell^{\alpha}\big\}}
\nonumber\\
\hspace*{-15mm}
&&\times\int_{-\pi}^{\pi}\prod_i\left[\frac{\mathrm d \hat k_i}{2\pi}~\rme^{\rmi\hat k_i k_i}\right]\rme^{-\rmi\lbrace\hat k_\ell+ \hat k_{\ell^\prime}\rbrace}
\nonumber\\
\hspace*{-15mm}
&&\hspace*{-10mm}\times
\exp\Big[\frac{c}{2N}\sum_{ij}\Big\lbrace\int\! \mathrm d J~ P(J)~  \rme^{-\rmi
J\sum_{\alpha}\big\{\hat{h}_i^{\alpha}
\sigma_j^{\alpha}+\hat{h}_j^{\alpha}
\sigma_i^{\alpha}\big\}-\rmi\lbrace\hat k_i+ \hat k_j\rbrace}\!-\!1\Big\rbrace+O(N^{0})\Big].
\label{eq2:All}
\end{eqnarray}
In the derivation of the above result we have used the integral representation of unity
 \begin{eqnarray}
1=\int\prod_{\alpha i} \mathrm d H_i^{\alpha}\delta
[H_i^{\alpha}- h_i (\sigmav^{\alpha})]\label{def2:Hdelta}
\end{eqnarray}
and the integral representation of the Kronecker $\delta$-functions
\begin{eqnarray}
\delta_{k_i,\sum_{j\neq i} c_{ij}}&=&\int_{-\pi}^{\pi}\frac{\mathrm d \hat k_i}{2\pi}\rme^{\rmi\hat k_i(k_i-\sum_{j\neq i} c_{ij})}.\label{def2:Kdelta}
\end{eqnarray}
In order to disentangle the $N$ degrees of freedom in equation (\ref{eq2:All}) we next define a replica density function
\begin{eqnarray}
\hspace{-5mm}P(\sigmav,\hat{\hv},\hat k ;\{\sigmav_i\},\{\hat{\hv}_i\},\{\hat k_i\})=\frac{1}{N}\sum_i
\delta_{\sigmav,\sigmav_i}\delta [\hat{\hv}-\hat{\hv}_i]\delta [\hat{k}-\hat{k}_i]\label{def2:P}
\end{eqnarray}
where $\sigmav=(\sigma_1,\ldots\sigma_n)$,
$\sigmav_i=(\sigma^1_i,\ldots\sigma^n_i)$ (similarly for the
replicated vectors $\hat{\hv}$, etc.), via insertion into equation (\ref{eq2:All}) of the following $\delta$-functional unity representation
\begin{eqnarray}
&&\hspace{-15mm}1=\int \prod_{\sigmav, \hat{\hv},\hat k}\mathrm d P(\sigmav ,
\hat{\hv},\hat k)~\delta \left[P(\sigmav , \hat{\hv},\hat k)-P(\sigmav ,
\hat{\hv},\hat k ;\{\sigmav_i\},\{\hat{\hv}_i\},\{\hat k_i\})\right]
\end{eqnarray}
which gives, with the short-hands $\bra g(J) \ket_J=\int\!\rmd J~P(J)g(J)$ and $\bx\cdot\by=\sum_\alpha x^\alpha y^\alpha$,
\begin{eqnarray}
\hspace*{-15mm}
&&\hspace{-15mm}\left\langle c_{\ell\ell^\prime}A_{\ell\ell^\prime}[s,s^\prime;h,h^\prime;\tilde s]\right\rangle_{\{c_{ij}J_{ij}\}}
\\
\hspace*{-15mm}
&=&\frac{1}{\Z}\frac{c}{N}\int\!\prod_{\tau h \alpha}\frac{\mathrm
d\hat{D}_{\alpha}(\tau,h
)}{2\pi/N}\int\! \prod_{\sigmav,
\hat{\hv},\hat k}\frac{\mathrm d \hat{P}(\sigmav , \hat{\hv},\hat k)\mathrm d
P (\sigmav , \hat{\hv},\hat k)}{2\pi/N}
\nonumber\\
\hspace*{-15mm}
&&\times\exp \Big[\rmi
N\sum_{\tau ,h ,\alpha}\hat{D}_{\alpha}(\tau ,h
)D(\tau ,h)+\rmi N\sum_{\sigmav}\int \mathrm d \hat{\hv}\mathrm d \hat k\hat{P}(\sigmav ,
\hat{\hv},\hat k)P (\sigmav , \hat{\hv},\hat k)
\nonumber\\
\hspace*{-15mm}
&&+\frac{1}{2}c N
\sum_{\sigmav,\sigmav^\prime}\int\! \mathrm d \hat{\hv}\mathrm d
\hat{\hv}^\prime \mathrm d \hat k \mathrm d \hat{k}^\prime P (\sigmav , \hat{\hv},\hat k)P (\sigmav^\prime ,
\hat{\hv}^\prime,\hat{k}^\prime)
\nonumber\\[-2mm]
\hspace*{-15mm}
&&\hspace*{40mm}\times\Big\langle \rme^{-\rmi J[\hat{\hv} .\sigmav^\prime+\hat{\hv}^{\prime} .\sigmav]-\rmi[\hat k+\hat{k}^\prime]}-1\Big\rangle_J+O(N^{0})\Big]
\nonumber\\
\hspace*{-15mm}
&&\times\sum_{\sigmav^1}\ldots\sum_{\sigmav^n}\int\!\prod_{i}\Big[\frac{
\mathrm d \Hv_i\mathrm d
\hat{\hv}_i}{2\pi}\Big]\int_{-\pi}^{\pi}\prod_i\Big[\frac{\mathrm d \hat k_i}{2\pi}~\rme^{\rmi\hat k_i k_i}\Big]
\exp\Big[\rmi\sum_{i}\hat{\hv}_i
.\left\{\Hv_i-\thetav\right\}\Big]
\nonumber\\
\hspace*{-15mm}
&&\times\exp\Big[ - \rmi\sum_{\tau ,h,\alpha}
\hat{D}_{\alpha}(\tau ,h
)\sum_i\delta_{\tau ,\sigma_i^{\alpha}}\delta\left[h
-H_i^{\alpha}\right]-\rmi\sum_i\hat{P}(\sigmav_i , \hat{\hv}_i, \hat k_i)\Big]
\nonumber\\
\hspace*{-15mm}
&&\times\delta_{s^\prime\!,\sigma_\ell^1}\delta_{s,\sigma_{\ell^\prime}^1}\delta
[h^\prime\!-H_\ell^1]
 \Big\langle\delta [ h \!-\!H_{\ell^\prime}^1\!+\!2J\tilde
s]\rme^{-\rmi J[\hat{\hv}_\ell .\sigmav_{\ell^\prime}+
\hat{\hv}_{\ell^\prime} .\sigmav_\ell]}\Big\rangle_{\!J}
 \rme^{-\rmi\lbrace\hat k_\ell+ \hat k_{\ell^\prime}\rbrace}.\nonumber
\end{eqnarray}
Inserting the above result into the sum (\ref{def2:A}), followed by further manipulations (see \ref{section2:A} for details), leads us to the path integral
\begin{eqnarray}
&&\hspace{-25mm}A[s,s^\prime;h,
h^\prime;\tilde s] =\lim_{N\rightarrow\infty}\lim_{n\rightarrow
0}\frac{1}{\Z}\left[\frac{1}{2\pi}\right]^{N}\int\!\{\mathrm d P \mathrm d\hat P \mathrm d\hat
D\}~\rme^{N\Psi[\{P ,\hat P ,\hat{D}\}]+ O(N^0)}\nonumber
\\
&&\times\left\{\sum_{k,k^\prime\geq0}P_c(k)P_c(k^\prime)\sum_{\sigmav,\sigmav^\prime}\int
\mathrm d \Hv\mathrm d \Hv^\prime\mathrm d
\hat{\hv}\mathrm d
\hat{\hv}^\prime
\right.
\nonumber\\
&&
\left. \times M [\Hv,\hat{\hv},\sigmav\vert k\!-\!1,\theta]~M [\Hv^\prime,\hat{\hv}^\prime,\sigmav^\prime\vert k^\prime\!-\!1,\theta]
\right.
\nonumber\\
&&\left.
\times\delta_{s^\prime ,\sigma_1}\delta_{s,\sigma^\prime_1}\delta[h^\prime\!-\!H_1]
 \left\langle\delta [h\!-\!H^\prime_1\!+\!2J\tilde
s]~\rme^{-\rmi J[\hat{\hv} .\sigmav^\prime+
\hat{\hv}^\prime.\sigmav]}\right\rangle_J
\right.
\label{eq2:Aintegral}
\\
&&\left.\hspace*{-10mm}
\times\Big[\sum_{\sigmav,\sigmav^\prime}\int\!
\mathrm d \Hv\mathrm d \Hv^\prime\mathrm d
\hat{\hv}\mathrm d
\hat{\hv}^\prime M [\Hv,\hat{\hv},\sigmav\vert k,\theta]~M [\Hv^\prime\!,\hat{\hv}^\prime\!,\sigmav^\prime\vert k^\prime\!,\theta]\Big]^{-1}
\!+~O(N^{-1})\nonumber\bigroom\right\}
\nonumber
\end{eqnarray}
where we use the following definitions:
\begin{eqnarray}
&&\hspace{-25mm}\Psi[\{P ,\hat{P},\hat{D}\}]=\rmi\sum_{\tau ,h,\alpha}\Delta
h
\hat{D}_{\alpha}(\tau ,h
)D(\tau ,h
) +\rmi\sum_{\sigmav}\int \mathrm d \hat{\hv}\mathrm d \hat k\hat{P}(\sigmav ,
\hat{\hv},\hat k)P (\sigmav , \hat{\hv},\hat k)
\nonumber
\\
&&\hspace*{-10mm}
+\frac{1}{2}c
\sum_{\sigmav,\sigmav^\prime}\int\! \mathrm d \hat{\hv}\mathrm d
\hat{\hv}^\prime \mathrm d \hat k \mathrm d \hat{k}^\prime P (\sigmav,\hat{\hv},\hat k)P (\sigmav^\prime\!,
\hat{\hv}^\prime\!,\hat{k}^\prime)\Big\langle \rme^{-\rmi J[\hat{\hv}.\sigmav^\prime+\hat{\hv}^{\prime} .\sigmav]-\rmi[\hat k+\hat{k}^\prime]}\!-\!1\Big\rangle_J
\nonumber\\
&&+\sum_{k\geq0}P_c(k)\log \sum_{\sigmav}\int\mathrm d \Hv\mathrm d
\hat{\hv}M [\Hv,\hat{\hv},\sigmav\vert k,\theta]
\label{def2:Saddle}
\end{eqnarray}
and
\begin{eqnarray}
&&\hspace{-15mm}M [\Hv,\hat{\hv},\sigmav\vert k\!-\!m,\theta]=\int_{-\pi}^{\pi}\!\mathrm d \hat k~\rme^{-\rmi\hat k m}~M [\Hv,\hat{\hv},\sigmav\vert k,\hat k,\theta]\label{def2:M}\\
&&\hspace{-15mm}M [\Hv,\hat{\hv},\sigmav\vert k,\hat k,\theta]=\frac{1}{2\pi}~\rme^{\rmi\hat{\hv}
.\{\Hv-\thetav\}- \rmi\sum_{\tau ,h,\alpha}\Delta
h
\hat{D}_{\alpha}(\tau ,h
)\delta_{\tau ,\sigma_{\alpha}}\delta\left[h
-H_{\alpha}\right]+\rmi\hat k k-\rmi\hat{P}(\sigmav , \hat{\hv},\hat k)}\nonumber
\end{eqnarray}
with $m\in\mathbb{Z}$. Finally, we change the order of the limits $N\rightarrow\infty$ and $n\rightarrow0$ in (\ref{eq2:Aintegral}) and, with the help of the normalization identity $\sum_{s,s^\prime}\int\!\mathrm d h \mathrm d h^\prime A[s,s^\prime;h,h^\prime;\tilde s] =1$, we compute (\ref{eq2:Aintegral}) by steepest descent, which gives
\begin{eqnarray}
&&\hspace{-25mm}A[s,s^\prime;h,
h^\prime;\tilde s] =\lim_{n\rightarrow
0}\frac{1}{Z_A}\sum_{k,k^\prime}P_c(k)P_c(k^\prime)\sum_{\sigmav,\sigmav^\prime}\int\!\mathrm d \Hv\mathrm d \Hv^\prime\mathrm d
\hat{\hv}\mathrm d
\hat{\hv}^\prime
\nonumber\\
&&\times M [\Hv,\hat{\hv},\sigmav\vert k\!-\!1,\theta]~M [\Hv^\prime,\hat{\hv}^\prime,\sigmav^\prime\vert k^\prime\!-\!1,\theta]\nonumber\\
&&\times\delta_{s^\prime ,\sigma_1}\delta_{s,\sigma_1^\prime}\delta[h^\prime  -H_1]
 \Big\langle\delta [ h -H_1^\prime+2J\tilde
s]\rme^{-\rmi J[\hat{\hv} .\sigmav^\prime+
\hat{\hv}^{\prime} .\sigmav]}\Big\rangle_{\!J}\nonumber\\
&&\times \Big[\sum_{\sigmav,\sigmav^\prime}\int\!\mathrm d \Hv\mathrm d \Hv^\prime\mathrm d
\hat{\hv}\mathrm d
\hat{\hv}^\prime M [\Hv,\hat{\hv},\sigmav\vert k,\theta]~M [\Hv^\prime,\hat{\hv}^\prime,\sigmav^\prime\vert k^\prime,\theta]
\Big]^{-1}
\label{eq2:A}
\end{eqnarray}
where $Z_A$ is a constant that ensures the proper normalization of $A$.
The order parameters $\{P,\hat P,\hat D\}$ are determined by extremization of the functional $\Psi$ in (\ref{def2:Saddle}), which leads us to four functional saddle-point equations
\begin{eqnarray}
\hspace*{-15mm}
D(\sigma ,h)&=&\sum_{k\geq0}P_c(k)\frac{\sum_{\sigmav}\int\mathrm d \Hv\mathrm d
\hat{\hv}~M [\Hv,\hat{\hv},\sigmav\vert k,\theta]~\delta_{\sigma,\sigma_\gamma}\delta(h-H_{\gamma})}{\sum_{\sigmav}\int\mathrm d \Hv\mathrm d
\hat{\hv}~M [\Hv,\hat{\hv},\sigmav\vert k,\theta]}~\label{eq2:D}
\\
\hspace*{-15mm}
P (\sigmav, \hat{\hv},\hat k)&=&\sum_{k\geq0}P_c(k)\frac{\int\!\mathrm d \Hv~M [\Hv,\hat{\hv},\sigmav\vert k,\hat k,\theta]}{\sum_{\sigmav}\int\!\mathrm d \Hv\mathrm d
\hat{\hv}~M [\Hv,\hat{\hv},\sigmav\vert k,\theta]}\label{eq2:P}
\\
\hspace*{-15mm}
\hat{P}(\sigmav ,\hat{\hv}, \hat k)&=&\rmi ~c~ Q(\sigmav ,\hat{\hv}, \hat k)
\label{eq2:hatP}
\\[2mm]
\hspace*{-15mm}
Q(\sigmav ,\hat{\hv}, \hat k)&=&\sum_{\sigmav^\prime}\int\!\mathrm d
\hat{\hv}^\prime \mathrm d \hat{k}^\prime P (\sigmav^\prime ,
\hat{\hv}^\prime,\hat{k}^\prime)\Big\langle \rme^{-\rmi J[\hat{\hv} .\sigmav^\prime+\hat{\hv}^{\prime} .\sigmav]-\rmi[\hat k+\hat{k}^\prime]}-1\Big\rangle_{\!J}
\label{eq2:Q}
\end{eqnarray}
The relations (\ref{eq2:hatP},\ref{eq2:Q}) allow us to relate the order parameter $P(\sigmav, \hat{\hv},\hat k)$ to its conjugate $\hat{P}(\sigmav ,\hat{\hv}, \hat k)$, and thereby remove the latter from the function $M$ (\ref{def2:M}). Furthermore, assuming that the function $\hat D_\alpha(s,h)$ is well behaved in the sense that $\sum_h\Delta h~ D_\alpha(s,h)g(h)\rightarrow\int\mathrm\! d h~D_\alpha(s,h)g(h)$ for $\Delta h\to 0$, we have
\begin{eqnarray}
\hspace*{-5mm}
M [\Hv,\hat{\hv},\sigmav\vert k,\hat k,\theta]&=&\frac{1}{2\pi}~\rme^{\rmi\hat{\hv}
.\{\Hv-\thetav\}-\rmi \sum_{\alpha}\hat{D}_{\alpha}(\sigma_\alpha ,H_{\alpha})+\rmi\hat k k+c Q(\sigmav , \hat{\hv},\hat k)}\label{eq2:M}
\end{eqnarray}
in definition (\ref{def2:M}).

The conjugate parameters $\hat{D}_{\alpha}(\sigma,h)$ and $\hat{P}(\sigmav, \hat{\hv},\hat k)$ in our replica theory play the role of Lagrange multipliers enforcing the normalization of the joint spin-field distribution $D(\sigma,h)$ and of the replica density function $P (\sigmav, \hat{\hv},\hat k)$. The physical meaning of the density $P (\sigmav, \hat{\hv},\hat k)$ is not yet clear due to the presence of the vector $\hat{\hv}$ and the parameter $\hat k$. However, we note that in our theory only the Fourier transforms  $\int\!\mathrm d\hat{\hv}~\rme^{-\rmi\xv.\hat{\hv}}\int_{-\pi}^{\pi}\mathrm d \hat k~\rme^{-\rmi\hat k}P(\sigmav, \hat{\hv},\hat k)$ of this function are relevant, where $\xv\in\mathbb{R}^{n}$.
\subsection{Equilibrium}\label{section2:Equilibr}
 In this section we show that the equilibrium solution of the model (\ref{def2:E}) is also a stationary solution of our dynamic equation (\ref{eq2:diffusion}). This is done in two steps.
First, we show that the equilibrium replica theory of the model under study is a special case of our dynamical replica theory. In order to do this,  we make an ansatz as in \cite{DRTinf2}:
\begin{eqnarray}
\rme^{-\rmi\sum_\alpha\hat{D}_{\alpha}(\sigma_{\alpha},H_{\alpha})}=\rme^{\frac{1}{2}\beta\sum_\alpha\sigma_{\alpha}\lbrace H_{\alpha}+\theta\rbrace}\label{eq2:DconjEq}
\end{eqnarray}
 and evaluate the Fourier transform of the replica density (\ref{eq2:P}), viz.
\begin{eqnarray}\int\mathrm d\hat{\hv}~\rme^{-\rmi\xv.\hat{\hv}}\int_{-\pi}^{\pi}\!\mathrm d \hat k~\rme^{-\rmi\hat k m}P(\sigmav, \hat{\hv},\hat k)\label{def2:Pft}\end{eqnarray}
for $\xv\in\mathbb{R}^{n}$ and $m\in\mathbb{Z}$. Using the saddle-point equation (\ref{eq2:P}) for the order parameter function $P(\sigmav, \hat{\hv},\hat k)$, combined with the Fourier transform of the function $M$ (see \ref{section2:M} for details),
\begin{eqnarray}
&&\hspace{-25mm}\int\mathrm d\hat{\hv}~\rme^{-\rmi\xv.\hat{\hv}}\int_{-\pi}^{\pi}\mathrm d \hat k~\rme^{-\rmi\hat k m}M [\Hv,\hat{\hv},\sigmav\vert k,\hat k,\theta]
\nonumber\\
&&\hspace{-10mm}=\frac{\rme^{-c}c^{k-m}}{(k-m)!}\prod_{\ell=1}^{k-m}\left[\sum_{\sigmav_\ell}\int\mathrm d
\hat{\hv}_\ell \mathrm d J_\ell P(J_\ell)\int_{-\pi}^{\pi}\mathrm d \hat{k}_\ell P (\sigmav_\ell ,
\hat{\hv}_\ell,\hat{k}_\ell)\rme^{-\rmi\hat{k}_\ell}\rme^{-\rmi J_\ell\hat{\hv}_\ell .\sigmav}\right]\nonumber\\
&&\hspace{-5mm}\times(2\pi)^n~\delta\Big[\Hv-\sum_\ell J_\ell\sigmav_\ell-\thetav-\xv\Big]~\rme^{- \rmi\sum_{\alpha}\hat{D}_{\alpha}(\sigma_\alpha ,H_{\alpha})}
\label{eq2:Mft}
\end{eqnarray}
leads us to the desired result for (\ref{def2:Pft})
\begin{eqnarray}
&&\hspace{-25mm}\int\mathrm d\hat{\hv}~\rme^{-\rmi\xv.\hat{\hv}}\int_{-\pi}^{\pi}\mathrm d \hat k~\rme^{-\rmi\hat k m}P(\sigmav, \hat{\hv},\hat k)\label{eq2:Pft}\\
&=&\sum_{k\geq m}P_c(k)\frac{1}{M_k}\frac{k!}{(k-m)!}c^{-m}~\rme^{\frac{1}{2}\beta\sigmav.\{\sum_\ell J_\ell\sigmav_\ell+2\thetav+\xv\}}\nonumber\\
&&\times\prod_{\ell=1}^{k-m}\left[\sum_{\sigmav_\ell}\int\mathrm d
\hat{\hv}_\ell \mathrm d J_\ell P(J_\ell)\int_{-\pi}^{\pi}\mathrm d \hat{k}_\ell P (\sigmav_\ell ,
\hat{\hv}_\ell,\hat{k}_\ell)\rme^{-\rmi\hat{k}_\ell}\rme^{-\rmi J_\ell\hat{\hv}_\ell .\sigmav}\right]\nonumber
\end{eqnarray}
given the ansatz (\ref{eq2:DconjEq}), where we define
\begin{eqnarray}
&&\hspace{-15mm}M_k=\sum_{\sigmav}\prod_{\ell=1}^{k}\left[\sum_{\sigmav_\ell}\int\mathrm d
\hat{\hv}_\ell \mathrm d J_\ell P(J_\ell)\int_{-\pi}^{\pi}\mathrm d \hat{k}_\ell P (\sigmav_\ell ,
\hat{\hv}_\ell,\hat{k}_\ell)\rme^{-\rmi\hat{k}_\ell}
\rme^{-\rmi J_\ell\hat{\hv}_\ell .\sigmav}\right]~~~~~\label{def2:Mk}\\
&&\times\rme^{\frac{1}{2}\beta\sigmav.\{\sum_\ell J_\ell\sigmav_\ell+2\thetav\}}.\nonumber
\end{eqnarray}
Solving equation (\ref{eq2:Pft}) for $m=1$ yields a very useful equality,
\begin{eqnarray}
&&\hspace{-15mm}\int\!\mathrm d\hat{\hv}~\rme^{-\rmi\xv.\hat{\hv}}\int_{-\pi}^{\pi}\mathrm d \hat k~\rme^{-\rmi\hat k}P(\sigmav, \hat{\hv},\hat k)=\int_{-\pi}^{\pi}\!\mathrm d \hat k~\rme^{-\rmi\hat k}P(\sigmav,\hat k)~\rme^{\frac{1}{2}\beta\sigmav.\xv}\label{eq2:FTeq}
\end{eqnarray}
which allows us to compute the integrals over $\{\hat{\hv}_\ell\}$ in (\ref{eq2:Pft}), giving
\begin{eqnarray}
&&\hspace{-25mm}\int\mathrm d\hat{\hv}~\rme^{-\rmi\xv.\hat{\hv}}\int_{-\pi}^{\pi}\mathrm d \hat k~\rme^{-\rmi\hat k m}P(\sigmav, \hat{\hv},\hat k)\label{eq2:Pft1}\\
&&\hspace{-15mm}=\sum_{k\geq m}P_c(k)\frac{k!}{(k-m)!}c^{-m}\frac{\left[\sum_{\sigmav^\prime} \tilde P (\sigmav^\prime )\int\mathrm d J P(J)~\rme^{\beta J\sigmav .\sigmav^\prime}\right]^{k-m}\rme^{\beta\sigmav.\thetav}~\rme^{\frac{1}{2}\beta\sigmav.\xv}}{\sum_{\sigmav^{\prime\prime}}\left[\sum_{\sigmav^{\prime\prime\prime}} \tilde P (\sigmav^{\prime\prime\prime})\int\mathrm d J P(J)~\rme^{\beta J\sigmav^{\prime\prime} .\sigmav^{\prime\prime\prime}}\right]^{k}\rme^{\beta\sigmav^{\prime\prime}.\thetav}}\nonumber
\end{eqnarray}
with the  shorthand $\tilde P(\sigmav)=\int_{-\pi}^{\pi}\mathrm d \hat k~\rme^{-\rmi\hat k}P(\sigmav,\hat k)$. Now for $\xv=(0,\ldots,0)$ and $m=1$ our equation (\ref{eq2:Pft1}) acquires the form
\begin{eqnarray}
&&\hspace{-15mm}\tilde P(\sigmav)=\sum_{k\geq1}\frac{P_c(k)k}{c}\frac{\left[\sum_{\sigmav^\prime} \tilde P (\sigmav^\prime)\int\mathrm d J P(J)~\rme^{\beta J\sigmav .\sigmav^\prime}\right]^{k-1}\rme^{\beta\sigmav.\thetav}}{\sum_{\sigmav^{\prime\prime}}\left[\sum_{\sigmav^{\prime\prime\prime}} \tilde P (\sigmav^{\prime\prime\prime})\int\mathrm d J P(J)~\rme^{\beta J\sigmav^{\prime\prime} .\sigmav^{\prime\prime\prime}}\right]^{k}\rme^{\beta\sigmav^{\prime\prime}.\thetav}}\label{eq2:Peq}
\end{eqnarray}
which is exactly the order parameter equation as found in equilibrium \cite{FCNN}.

The second part of our proof consists in showing that the ansatz (\ref{eq2:DconjEq}) also leads to a stationary solution of our present dynamic equation (\ref{eq2:diffusion}). For this purpose we compute the saddle-point equations for the joint spin-field probability distributions (\ref{eq2:A}) and (\ref{eq2:D}), given our ansatz. The result of this calculation (see \ref{section2:EquilibrCalc} for details) allows us to write these two equations in the form
\begin{eqnarray}
\hspace*{-10mm}
D(s,h)&=&\rme^{\beta s h}\Phi[h]\label{eq2:Deq}
\\
\hspace*{-10mm}
A[s,s^\prime ;h,
h^\prime;\tilde s]&=&\Big\langle \rme^{\beta s(h + 2J\tilde s)+\beta s^\prime  h^\prime -\beta Jss^\prime }\Lambda[h\!+\!2J\tilde s\!-\!Js^\prime \!;h^\prime \!-\!Js]\Big\rangle_{\!J}
\label{eq2:Aeq}
\end{eqnarray}
respectively, where all complicated terms dependent on the replicas are contained in the two functions $\Phi$ and $\Lambda$ (which are defined in
  \ref{section2:EquilibrCalc}). Inserting (\ref{eq2:Deq}) and (\ref{eq2:Aeq}) into the right-hand side of the dynamic equation (\ref{eq2:diffusion}), followed by further manipulations (see \ref{section2:FixedPoints}), leads us to the equality $\frac{\partial}{\partial t}D(s,h)=0$ for all $s\in\{-1,1\}$ and $h\in\mathbb{R}$, as claimed.
Thus, the equilibrium solution of the model (\ref{def2:E}) indeed defines a stationary state of the dynamics (\ref{eq2:diffusion}).
\subsection{Replica symmetry}\label{section2:RS}
In order to take the $n\rightarrow0$ limit in our equations (\ref{eq2:A})-(\ref{eq2:P}) we assume replica symmetry (RS). For the conjugate order parameters $\hat{D}_{\alpha}(s,H)$, which are  depend only on a single replica index and expected to be imaginary, this translates into
\begin{eqnarray}
\hat{D}_{\alpha}(s,H)&=&\rmi\log d(s,H).\label{def2:Drs}
\end{eqnarray}
The replica density $P(\sigmav, \hat{\hv},\hat k)$ depends on one discrete vector $\sigmav$ and one continuous vector $\hat{\hv}$ in replica space. The parameter $\hat k$ is a scalar variable coupled to the vertex degree $k$, which is a random variable. Replica symmetry demands that the order parameter $P(\sigmav, \hat{\hv},\hat k)$ is invariant under permutation of the replica indices, for any value of $\hat k$, which implies \cite{SoftSpins,thesis} that it is of the general form
\begin{eqnarray}
P_{RS}(\sigmav, \hat{\hv},\hat k)&=&\int\! \left\lbrace \mathrm d
P\right\rbrace~ W[\{P\};\hat k]\prod_{\alpha=1}^n
P(\sigma_{\alpha},\hat{h}_{\alpha})\label{def2:Prs}
\end{eqnarray}
where $W[\{P\};\hat k]$ is a functional distribution, which must be normalized according to $\int\! \left\lbrace \mathrm d
P\right\rbrace\int_{-\pi}^{\pi}\mathrm d \hat k~ W[\{P\};\hat k]=1$. It turns out that also the Fourier transform $\int_{-\pi}^{\pi}\!\mathrm d \hat k~W[\{P\};\hat k]~\rme^{-\rmi\hat k}$ of this functional distribution is normalized\footnote{This can be shown by substituting (\ref{def2:Prs}) into the function $\Psi$  (\ref{def2:Saddle}), followed by expanding this function for small $n$. The desired result $\int\!\lbrace \mathrm d
P\rbrace\int_{-\pi}^{\pi}\mathrm d \hat k~ W[\{P\};\hat k]~\rme^{-\rmi\hat k}=1 $ then follows from solving the saddle-point equations for the $O(n^0)$ part of $\Psi$.}, which is very convenient for our further calculations.

The RS ansatz (\ref{def2:Drs},\ref{def2:Prs}) allows us to take the replica limit $n\rightarrow0$ in equations (\ref{eq2:A})-(\ref{eq2:P}). The Fourier transform $\int_{-\pi}^{\pi}\!\mathrm d \hat k~\rme^{-\rmi\hat k m}M[\Hv,\hat{\hv},\sigmav\vert k,\hat k,\theta]$, where $m\in\mathbb{Z}$, is the main ingredient of these equations. We can easily  compute its RS version using result (\ref{eq2:Mft}),  leading to
\begin{eqnarray}
\hspace*{-15mm}
&&\hspace*{-15mm}\int_{-\pi}^{\pi}\mathrm d \hat k~\rme^{-\rmi \hat k m}M_{RS} [\Hv,\hat{\hv},\sigmav\vert k,\hat k,\theta]
\label{eq2:Mrs}\\
\hspace*{-15mm}
&=&\frac{e^{-c}c^{k-m}}{(k-m)!}\int\!\prod_{\ell=1}^{k-m}\Big[\mathrm d J_\ell P(J_\ell) \left\lbrace \mathrm d
P_\ell\right\rbrace~ \int_{-\pi}^{\pi}\mathrm d \hat{k}_\ell W[\{P_\ell\};\hat{k}_\ell]~\rme^{-\rmi \hat{k}_\ell}\Big]
\nonumber\\
\hspace*{-15mm}
&&\times\prod_{\alpha=1}^n\Big\{ d(\sigma_\alpha,H_\alpha)\rme^{\rmi\hat{h}_\alpha
\{H_\alpha-\theta\}}\prod_{\ell=1}^{k-m}\Big[\sum_{\sigma_{\ell}^{\alpha}}\int\!\mathrm d
\hat{h}_{\ell}^{\alpha}
P_\ell(\sigma_{\ell}^{\alpha},\hat{h}_{\ell}^{\alpha})\rme^{-\rmi J_\ell[\hat{h}_{\ell}^{\alpha}\sigma_\alpha+\hat{h}_\alpha\sigma_{\ell}^{\alpha}]}\Big]\Big\}\nonumber.
\end{eqnarray}
Now we can use (\ref{eq2:Mrs}) and the saddle-point equation (\ref{eq2:P}) to solve for the functional distribution $W[\{P\};\hat k]$. However, it is clear from (\ref{eq2:Mrs}) that all the equations of our theory are dependent only on $\int_{-\pi}^{\pi}\mathrm d \hat k~W[\{P\};\hat k]~\rme^{-\rmi \hat k m}$, rather than on the distribution $W[\{P\};\hat{k}]$ itself. Thus for $m\in\mathbb{Z}$ we define
\begin{eqnarray}
W[\{P\}\vert m]=\int_{-\pi}^{\pi}\mathrm d \hat k~W[\{P\};\hat k]~\rme^{-\rmi \hat k m}\label{def2:Wm}
\end{eqnarray}
and compute this object (see \ref{section2:RScalc}), which leads us to the equation
\begin{eqnarray}
&\hspace*{-25mm}W[\{P\}\vert m]=\sum_{k\geq m}P_c(k)\frac{k!}{(k- m)!}c^{-m}\int\!\prod_{\ell=1}^{k-m}\left[\mathrm d J_\ell P(J_\ell) \left\lbrace \mathrm d
P_\ell\right\rbrace W[\{P_\ell\}\vert1]\right]\label{eq2:W}\\
&\hspace*{-23mm}\times\prod_{\sigma,\hat h}\delta\left[P(\sigma,\hat h)\!-\frac{\int\!\mathrm d H d(\sigma,H)\rme^{\rmi\hat{h}
[H-\theta]}\prod_{\ell=1}^{k\minus m}\Big[\sum_{\sigma_{\ell}}\int\!\mathrm d
\hat{h}_{\ell}
P_\ell(\sigma_{\ell},\hat{h}_{\ell})~\rme^{-\rmi J_\ell[\hat{h}_{\ell}\sigma+\hat{h}\sigma_{\ell}]}\Big]}{Z[\{P_1,\ldots,P_{k-m}\}]}\right]
\nonumber
\end{eqnarray}
where $m\in\{0,1\}$, and $Z[\ldots]$ is a normalization constant given by
\begin{eqnarray}
&\hspace*{-20mm}Z[\{P_1,\ldots,P_{k-m}\}]=&2\pi\sum_{\sigma^\prime}\prod_{\ell=1}^{k-m}\left[
\sum_{\sigma_{\ell}}\int\mathrm d \hat{h}_{\ell}
P_\ell(\sigma_{\ell},\hat{h}_{\ell})~\rme^{- \rmi J_\ell\hat{h}_{\ell}
\sigma^\prime}\right] d\big(\sigma^\prime,\sum_{\ell}
J_\ell\sigma_{\ell}\!+\!\theta\big)\nonumber\\
&&
\end{eqnarray}
It is easy to verify that for $m\in\{0,1\}$ the functional distribution (\ref{eq2:W})  is normalized for any vertex degree distribution $P_c(k)$, provided the latter satisfies $\sum_k P_c(k) k=c$.

 Next we compute the kernels (\ref{eq2:A}) and (\ref{eq2:D}) under the RS ansatz. This is done by using (\ref{eq2:Mrs}) in both, followed by the replica limit (see \ref{section2:RScalc}), giving
\begin{eqnarray}
\hspace*{-15mm}D(\sigma ,h)&=&\sum_{k\geq0}P_c(k)~\int\!\prod_{\ell=1}^{k}\left[\mathrm d J_\ell P(J_\ell) \left\lbrace \mathrm d
P_\ell\right\rbrace~ W[\{P_\ell\}\vert1]\right]\label{eq2:Drs}
\\
\hspace*{-15mm}
&&\times\frac{d(\sigma,h)\!\prod_{\ell=1}^{k}\left[
\sum_{\sigma_{\ell}}\int\!\mathrm d \hat{h}_{\ell}
P_\ell(\sigma_{\ell},\hat{h}_{\ell})\rme^{- i J_\ell\hat{h}_{\ell}
\sigma}\right] ~\delta(h-\!\sum_{\ell}
J_\ell\sigma_{\ell}\!-\!\theta)}{\sum_{\sigma}\!\prod_{\ell=1}^{k}\left[
\sum_{\sigma_{\ell}}\int\!\mathrm d \hat{h}_{\ell}
P_\ell(\sigma_{\ell},\hat{h}_{\ell})\rme^{- i J_\ell\hat{h}_{\ell}
\sigma}\right] d\big(\sigma,\!\sum_{\ell}
J_\ell\sigma_{\ell}\!+\!\theta\big)}
\nonumber
\end{eqnarray}
and
\begin{eqnarray}
&&\hspace{-25mm}A[s,s^\prime;h,
h^\prime;\tilde s] =\sum_{k,k^\prime\geq1}\frac{P_c(k)k}{c}\frac{P_c(k^\prime)k^\prime}{c}\int\!\mathrm d J~ P(J)
\label{eq2:Ars}\\
&&\hspace{-10mm}
\times
\Big[\int\!\prod_{\ell=1}^{k-1}\!\left[\mathrm d J_\ell P(J_\ell) \left\lbrace \mathrm d
P_\ell\right\rbrace W[\{P_\ell\}\vert1]\right]\Big]
\Big[\int\!\prod_{r=1}^{k^\prime-1}\!\left[\mathrm d J^{\prime}_r P(J^{\prime}_r) \left\lbrace \mathrm d
Q_r\right\rbrace W[\{Q_r\}\vert1]\right]\Big]\nonumber\\
&&\hspace{-3mm}\times\sum_{\sigma,\sigma^\prime}\prod_{\ell=1}^{k-1}\left[\sum_{\sigma_{\ell}}\int\mathrm d
\hat{h}_{\ell}
P_\ell(\sigma_{\ell},\hat{h}_{\ell})~\rme^{-\rmi J_\ell\hat{h}_{\ell}\sigma}\right]~d(s^\prime,h^\prime)\nonumber\\
&&\hspace{-3mm}\times\prod_{r=1}^{k^\prime-1}\left[\sum_{\sigma_{r}}\int\mathrm d
\hat{h}_{r}
Q_r(\sigma_{r},\hat{h}_{r})~\rme^{-\rmi J^{\prime}_r\hat{h}_{r}\sigma^{\prime}}\right]~d(s,h+2J\tilde
s)\nonumber\\
&&\hspace{-3mm}\times\delta_{s^\prime ,\sigma}\delta_{s,\sigma^\prime}\delta
[h^\prime\! -\sum_\ell J_\ell\sigma_\ell-\theta-J\sigma^\prime]
 \delta [ h -\sum_r J^{\prime}_r\sigma_r-\theta-J\sigma+2J\tilde
s]\nonumber\\
&&\hspace{-3mm}\times\Bigg\{\sum_{\sigma,\sigma^\prime}\prod_{\ell=1}^{k-1}\Big[\sum_{\sigma_{\ell}}\int\!\mathrm d
\hat{h}_{\ell}
P_\ell(\sigma_{\ell},\hat{h}_{\ell})~\rme^{-\rmi J_\ell\hat{h}_{\ell}\sigma}\Big]d(\sigma,\sum_\ell J_\ell\sigma_\ell+\theta+J\sigma^\prime)\nonumber\\
&&\hspace{-3mm}\times\prod_{r=1}^{k^\prime-1}\Big[\sum_{\sigma_{r}}\int\!\mathrm d
\hat{h}_{r}
Q_r(\sigma_{r},\hat{h}_{r})~\rme^{-\rmi J^{\prime}_r\hat{h}_{r}\sigma^{\prime}}\Big]d(\sigma^{\prime}\!,\sum_r J^{\prime}_r\sigma_r+\theta+J\sigma)\Bigg\}^{-1}\nonumber
\end{eqnarray}
where $\tilde s\in\{0,s^\prime\}$. Equations (\ref{eq2:W})-(\ref{eq2:Ars}) constitute the final analytic results of the RS theory in this section. The results of  a similar dynamical study \cite{MozeikaCoolen}, which was carried out for Poissonian graphs only, are easily recovered from the present  more general equations, by using the equality
\begin{eqnarray}
&&\sum_{k\geq m}P_c(k)~\frac{k!}{(k- m)!}~c^{-m}~a_{k-m}=\sum_{k\geq 0}P_c(k)~a_k,
\end{eqnarray}
(which holds for all $m\in\{0,1,\ldots,\}$ for the Poisson vertex degree distribution, i.e. when $P_c(k)={e^{-c}c^k}/{k!}$), throughout formulae (\ref{eq2:W})-(\ref{eq2:Ars}).

The solution of our dynamic equation (\ref{eq2:diffusion}) requires the computation of the kernel (\ref{eq2:Ars}) at every instance of time $t$. In order to compute this kernel we have to solve the saddle-point equations (\ref{eq2:W}) and (\ref{eq2:Drs}) for the functional distribution $W$ and the function $d(s,h)$, given the instantaneous values of the joint spin-field distribution $D(s,h)$ at time $t$. These equations are integro-functional, and analytic solution is generally ruled out. However, we can solve them numerically \cite{MozeikaCoolen} using the population dynamics \cite{BetheSG} algorithm. In order to apply this latter numerical method efficiently we may transform $W\rightarrow\tilde W$ and $\hat P(\sigma \vert x)\rightarrow \int\! \mathrm d \hat h
P(\sigma, \hat h)\rme^{\minus \rmi\hat h x} $ in equations (\ref{eq2:W})-(\ref{eq2:Ars}), according to\footnote{Here we assume that for $x\in\mathbb{R}$ the Fourier transforms $\int\! \mathrm d \hat h
P(\sigma, \hat h)\rme^{\minus \rmi\hat h x}$ are real-valued, which is certainly true in equilibrium (see section \ref{section2:Equilibr}) and is a self-consistent assumption for any time.}
\begin{eqnarray}
&&\hspace{-10mm}\tilde W[\{\hat P\}\vert1]=\int\left\lbrace \mathrm d
P\right\rbrace W[\{P\}\vert1]\prod_{\sigma x}
\delta\left[ \hat P(\sigma \vert x)- \int \mathrm d \hat h
P(\sigma, \hat h)~\rme^{\minus \rmi\hat h x}\right].\label{def2:Wft}
\end{eqnarray}
Upon substitution of (\ref{eq2:W}) into (\ref{def2:Wft}) we can easily derive the functional relation for (\ref{def2:Wft}), which is given by
\begin{eqnarray}
&&\hspace{-15mm}
\tilde W[\{\hat P\}\vert1]=\sum_{k\geq1}\frac{P_c(k) k}{c}\int\prod_{\ell=1}^{k-1}\left\lbrace \mathrm d J_\ell P(J_\ell)
\{\mathrm d \hat P_\ell\} \tilde W[\{\hat P_\ell\}\vert1]\right\rbrace\label{eq2:Wft}\\
&&\hspace{-0mm}\times\prod_{\sigma x}\delta\left[ \hat P(\sigma \vert x)-
\frac{\prod_{\ell=1}^{k-1}\left\lbrace \sum_{\sigma_{\ell}} \hat
P_\ell(\sigma_{\ell}\vert J_\ell\sigma)\right\rbrace d(\sigma,\sum_{\ell=1}^{k-1}
J_\ell\sigma_{\ell}\!+\! \theta\!+\! x)}{Z[\{\hat P_1,\ldots,\hat
P_{k-1}\}]}\right]\nonumber
\end{eqnarray}
where $Z[\{\hat P_1,\ldots,\hat P_{k-1}\}]=\sum_\sigma\prod_{\ell=1}^{k-1}\lbrace \sum_{\sigma_{\ell}} \hat
P_\ell(\sigma_{\ell}\vert J_\ell\sigma)\rbrace d(\sigma,\sum_{\ell=1}^{k-1}
J_\ell\sigma_{\ell}\!+\! \theta)$. The normalization of $\tilde W$ is seen to be built into this relation, however the functional arguments
$\hat P(\sigma \vert x)$ are only normalized for $x=0$.
%
\section{Dynamics of the diluted Ising ferromagnet in the Griffiths phase}\label{section2:Griffiths}
As an explicit application of the theory derived in previous sections, we now study the Glauber dynamics of the diluted Ising ferromagnet on the Bethe lattice.

\subsection{The model and its equilibrium properties}
We consider a model of an Ising ferromagnet characterized by the following Hamiltonian:
\begin{eqnarray}
H(\sigmav)=-\sum_{\langle i j\rangle} J_{ij}\sigma_i\sigma_j-\theta\sum_i
\sigma_i.\label{def2:Ediluted}
\end{eqnarray}
The sum is over all links of the Bethe lattice with connectivity $k$. The  bonds $J_{ij}$ are random and statistically independent: $J_{ij}=J$ with probability $p$ and $J_{ij}=0$ with probability $1-p$. The lattice contains only finite size clusters for $p<p_c$, where $p_c$ is the percolation threshold given by $p_c=1/(k-1)$ for the Bethe lattice \cite{FisherEssam}, whereas the giant cluster appears for $p>p_c$. The density of the finite clusters of bond-size $n$ is also known for the present model \cite{FisherEssam}, and asymptotically given by
\begin{eqnarray}
W_n(p,k)\sim n^{-\frac{5}{2}}\rme^{-n A(p,k)}~~~~~~(n\to\infty)\label{eq2:clusterDistr}
\end{eqnarray}
where
\begin{eqnarray}
A(p,k)=\ln\left[\frac{(k-2)^{k-2}}{(k-1)^{k-1}~p(1-p)^{k-2}}\right].
\end{eqnarray}
For $p=p_c$ we have $A(p,k)=0$ and the asymptotic form (\ref{eq2:clusterDistr}) of the density $W_n(p,k)$ is independent of $k$. The model (\ref{def2:Ediluted}) has paramagnetic and ferromagnetic phases, which are separated by the critical boundary \cite{Laumann}
\begin{eqnarray}
T_c(p)=J/\tanh^{-1}\!(p_c/p).\label{eq2:Tc}
\end{eqnarray}
The critical temperature of the undiluted Ising ferromagnet on the Bethe lattice is simply $T_c(1)$. Thus the Griffith phase of the model (\ref{def2:Ediluted}) is given by the range of temperatures $T_c(p)<T<T_c(1)$. The magnetization in the Griffiths phase and in the paramagnetic phase (i.e. for $T>T_c(1)$) vanishes, and without an external field (i.e. for $\theta=0$) the internal energy is given by
\begin{eqnarray}
\langle H(\sigmav)\rangle=-\frac{1}{2}pk\tanh(J/T)\label{eq2:Eeq}
\end{eqnarray}
  where the angular brackets define a thermal average (expression (\ref{eq2:Eeq}) is easily derived from the free energy in \cite{FCNN}). The presence of Griffiths singularities in the low temperature part of Griffiths region was demonstrated in \cite{Laumann} by studying the density of Yang-Lee zeroes \cite{YangLeeI,YangLeeII}. Moreover, the authors of \cite{Laumann} obtained an exact expansion for the cluster magnetizations, which was used in arguments by Harris \cite{Harris} for the site-diluted version of this problem, within the cavity approach. The presence of rare large clusters in the diluted Bethe lattice is responsible for the Griffiths effects in this model \cite{Laumann}. This singularity, however, is very weak ($\sim\rme^{-{\rm const}/\vert\theta\vert}$) and is difficult to observe in equilibrium \cite{Imry}.
In this paper we consider the Glauber dynamics of the diluted Ising ferromagnet (\ref{def2:Ediluted}) in the paramagnetic and Griffiths phases. To connect our dynamical theory, which was developed for random graphs parameterized by an arbitrary vertex degree distribution, with the equilibrium studies of this problem as carried out for Bethe lattices, we note that in the infinite system size limit $N\rightarrow\infty$ the random regular graphs defined by $P_c(k)=\delta_{k,c}$ asymptotically approach Bethe lattices \cite{Dorogovtsev}.

\subsection{Equations of the DRT for random regular graphs with dilution}\label{section2:DRTforRRG}
We can derive the order parameter equations (\ref{eq2:W},\ref{eq2:Drs},\ref{eq2:Ars}) for the diluted Ising ferromagnet simply  by inserting
into these three general equations the special choices $P_c(k)=\delta_{k,c}$ and $P(J_\ell)=p~\delta(J_\ell-J)+q~\delta(J_\ell)$, where $p\in[0,1]$ and $q=1-p$. Equation (\ref{eq2:W}) for the order parameter function $W$ is then simplified by summation over $k$. If we also replace $c\rightarrow k$
(since the non-diluted graph is regular), this leads us to
\begin{eqnarray}
&&\hspace{-25mm}W[\{P\}\vert 1]=\sum_{\tau_1,\ldots,\tau_{k-1}}P(\tau_1,\ldots,\tau_{k-1})\int\prod_{\ell=1}^{k-1}\left[ \left\lbrace \mathrm d
P_\ell\right\rbrace~  W[\{P_\ell\}\vert1]\right]\label{eq2:Wdil}\\
&&\hspace{-25mm}\times\prod_{\sigma,\hat h}\delta\left[P(\sigma,\hat h)-\frac{\int\mathrm d H ~d(\sigma,H)\rme^{\rmi\hat{h}
[H-\theta]}\prod_{\ell=1}^{k-1}\left[\sum_{\sigma_{\ell}}\int\mathrm d
\hat{h}_{\ell}
P_\ell(\sigma_{\ell},\hat{h}_{\ell})\rme^{-\rmi J\tau_\ell[\hat{h}_{\ell}\sigma+\hat{h}\sigma_{\ell}]}\right]}{Z[\{P_1,\ldots,P_{k-1}\}]}\right]\nonumber
\end{eqnarray}
where we define the probability function
\begin{eqnarray}
P(\tau_1,\ldots,\tau_k)=p^{\sum_{\ell=1}^k \delta_{\tau_\ell,1}}q^{k-\sum_{\ell=1}^k \delta_{\tau_\ell,1}}\label{def2:Ptau}
\end{eqnarray}
with the binary variable $\tau\in\{1,0\}$. We note that in equation (\ref{eq2:Wdil}) the terms with $\tau_\ell=0$ do not contribute, since the  $P_\ell(\sigma_{\ell},\hat{h}_{\ell})$ are normalized by definition. Finally, we transform $W[\{P\}\vert 1]\rightarrow \tilde W[\{\hat P\}\vert 1]$ in equation (\ref{eq2:Wdil}),  according to the definition
\begin{eqnarray}
&&\hspace{-15mm}\tilde W[\{\hat P\}\vert 1]=\int\left\lbrace \mathrm d P\right\rbrace W[\{P\}\vert 1]\prod_{\sigma\sigma^\prime}
\delta\left[ \hat P(\sigma \vert \sigma^\prime)-\int \mathrm d \hat h P(\sigma, \hat h)\rme^{-\rmi \hat h J \sigma^\prime}\right]~~~~~\label{def2:WdilFt}
\end{eqnarray}
where $\sigma,\sigma^\prime\in\{-1,1\}$, which leads us to an equation for the functional distribution of Fourier transforms:
\begin{eqnarray}
&&\hspace{-25mm}\tilde W[\{\hat P\}\vert 1]
=\sum_{\tau_1,\ldots,\tau_{k-1}}P(\tau_1,\ldots,\tau_{k-1})\int\!\prod_{\ell=1}^{k-1}\left[ \left\lbrace \mathrm d
\hat P_\ell\right\rbrace~  \tilde W[\{\hat P_\ell\}\vert1]\right]
\label{eq2:WdilFt}\\
&&\hspace{-12mm}\times\prod_{\sigma\sigma^\prime}\delta\left[ \hat P(\sigma \vert \sigma^\prime)-\frac{\prod_{\ell=1}^{k-1}\left\lbrace \sum_{\sigma_{\ell}}
\hat P_\ell(\sigma_{\ell}\vert\tau_\ell\sigma)\right\rbrace d(\sigma,J\sum_{\ell=1}^{k-1} \tau_\ell\sigma_{\ell}+\theta+J\sigma^\prime)}{\sum_{\sigma^{\prime\prime}}\prod_{\ell=1}^{k-1}\left\lbrace \sum_{\sigma_{\ell}}
\hat P_\ell(\sigma_{\ell}\vert\tau_\ell\sigma^{\prime\prime})\right\rbrace d(\sigma^{\prime\prime},J\sum_{\ell=1}^{k-1} \tau_\ell\sigma_{\ell}+\theta)}\right]\nonumber.
\end{eqnarray}
We can in fact get rid of the $\tau_\ell$ variables entirely, which gives us an alternative representation of the equation above
\begin{eqnarray}
&&\hspace{-25mm}\tilde W[\{\hat P\}\vert 1]
=\sum_{k^\prime=0}^{k-1}B_{k-1}(k^\prime)\int\!\prod_{\ell=1}^{k^\prime}\left[ \left\lbrace \mathrm d
\hat P_\ell\right\rbrace~  \tilde W[\{\hat P_\ell\}\vert1]\right]
\label{eq2:WdilFtbin}\\
&&\times\prod_{\sigma\sigma^\prime}\delta\left[ \hat P(\sigma \vert \sigma^\prime)-\frac{\prod_{\ell=1}^{k^\prime}\left\lbrace \sum_{\sigma_{\ell}}
\hat P_\ell(\sigma_{\ell}\vert\sigma)\right\rbrace d(\sigma,J\sum_{\ell=1}^{k^\prime} \sigma_{\ell}\!+\!\theta\!+\!J\sigma^\prime)}{\sum_{\sigma^{\prime\prime}}\prod_{\ell=1}^{k^\prime}\left\lbrace \sum_{\sigma_{\ell}}
\hat P_\ell(\sigma_{\ell}\vert\sigma^{\prime\prime})\right\rbrace d(\sigma^{\prime\prime},J\sum_{\ell=1}^{k^\prime} \sigma_{\ell}\!+\!\theta)}\right]\nonumber
\end{eqnarray}
where $B_{k-1}(k^\prime)$ is the binomial distribution
\begin{eqnarray}
&&B_{k-1}(k^\prime)=\left (\!\begin{array}{c}k\!-\!1\\k^\prime\end{array}
\!\right )p^{k^\prime}q^{k-1-k^\prime}\label{def2:binomial}.
\end{eqnarray}
This result reflects the fact that the distribution of the vertex degrees in the random regular graph of degree $k$ with the bond-dilution is indeed the binomial $B_{k}(k^\prime)$. The fields (\ref{eq2:field}) for the model (\ref{def2:Ediluted}) take the values $Jn+\theta$, where $n\in\{-k,\ldots,k\}$, allowing us to write the joint spin-field probability distributions (\ref{eq2:Drs}) and (\ref{eq2:Ars}) in the form
\begin{eqnarray}
&&\hspace{-15mm}D(s,h)=\sum_{n=-k}^{k}P(s,n)~\delta (h- J n-
\theta)\label{def2:Psn}
\\
&&\hspace{-15mm}A[s,s^\prime;h,h^\prime;\tilde s]=\sum_{n=-k+1}^{k-1}\sum_{n^\prime=-k+1}^{k-1}~\Big\langle A[s,s^\prime
;n,n^\prime\vert\tau]~\delta [h^{\prime} -Jn^{\prime}-\theta-J\tau s]
\nonumber
\\
&&\hspace*{20mm}\times
\delta [ h+2J\tau\tilde s -J n-\theta-J\tau s^{\prime}]\Big\rangle_\tau
\label{def2:Asn}
\end{eqnarray}
where $\langle\ldots\rangle_\tau=\sum_\tau~P(\tau)$, with $P(\tau)$ defined in (\ref{def2:Ptau}), and%
\begin{eqnarray}
&&\hspace{-15mm}P(s,n)=\sum_{\tau_1,\ldots,\tau_{k}}P(\tau_1,\ldots,\tau_{k})\int\!\prod_{\ell=1}^{k}\left[ \left\lbrace \mathrm d \hat
P_\ell\right\rbrace~  \tilde W[\{\hat P_\ell\}\vert 1]\right]\label{eq2:Psn}\\
&&\times\frac{d(s,Jn+\theta)\!\prod_{\ell=1}^{k}\left[
\sum_{\sigma_{\ell}}
\hat P_\ell(\sigma_{\ell}\vert \tau_\ell s)\right] ~\delta_{n;\sum_{\ell=1}^k
\tau_\ell\sigma_{\ell}}}{\sum_{\sigma}\!\prod_{\ell=1}^{k}\left[
\sum_{\sigma_{\ell}}\hat P_\ell(\sigma_{\ell}\vert \tau_\ell\sigma)\right] d\big(\sigma,\!J\sum_{\ell}
\tau_\ell\sigma_{\ell}\!+\!\theta\big)}\nonumber
\end{eqnarray}%
\begin{eqnarray}
&&\hspace{-15mm}A[s,s^\prime;n,
n^\prime\vert\tau] =\sum_{\tau_1,\ldots,\tau_{k-1}}P(\tau_1,\ldots,\tau_{k-1})\int\!\prod_{\ell=1}^{k-1}\left[\lbrace \mathrm d
\hat P_\ell\rbrace  \tilde W[\{\hat P_\ell\}\vert 1]~\right]\label{eq2:Asn}\\
&&\hspace{10mm}\times\sum_{\tau^\prime_1,\ldots,\tau^\prime_{k-1}}P(\tau^\prime_1,\ldots,\tau^\prime_{k-1})\int\!\prod_{r=1}^{k-1}\left[\lbrace \mathrm d
\hat Q_r\rbrace \tilde W[\{\hat Q_r\}\vert 1]~\right]\nonumber\\
&&\hspace{10mm}\times\prod_{\ell=1}^{k-1}\left[\sum_{\sigma_{\ell}}
\hat P_\ell(\sigma_{\ell}\vert \tau_\ell s^\prime)\right]~d(s^\prime,Jn^\prime+\theta+J\tau s)~\delta_{n^\prime;\sum_{\ell=1}^{k-1} \tau_\ell\sigma_\ell}\nonumber\\
&&\hspace{10mm}\times\prod_{r=1}^{k-1}\left[\sum_{\sigma_{r}}
\hat Q_r(\sigma_{r}\vert \tau^\prime_r s)\right]~d(s,Jn+\theta+J\tau s^\prime)~\delta_{n;\sum_{r=1}^{k-1} \tau^{\prime}_r\sigma_r}\nonumber\\
%
&&\hspace{7mm}\times\Bigg\{\sum_{\sigma,\sigma^\prime}\prod_{\ell=1}^{k-1}\left[\sum_{\sigma_{\ell}}
\hat P_\ell(\sigma_{\ell}\vert \tau_\ell \sigma)\right]~d(\sigma,J\sum_{\ell=1}^{k-1} \tau_\ell\sigma_\ell+\theta+J\tau\sigma^\prime)\nonumber\\
&&\hspace{15mm}\times\prod_{r=1}^{k-1}\left[\sum_{\sigma_{r}}
\hat Q_r(\sigma_{r}\vert \tau^\prime_r \sigma^\prime)\right]~d(\sigma^{\prime},J\sum_{r=1}^{k-1} \tau^{\prime}_r\sigma_r+\theta+J\tau\sigma)\Bigg\}^{-1}\nonumber
\end{eqnarray}
where in deriving probability distributions over the integer fields (\ref{eq2:Psn}) and (\ref{eq2:Asn}) we followed the steps leading to (\ref{eq2:WdilFt}). It is easy to show, using equation (\ref{eq2:WdilFt}), that the distribution $P(s,n)$ is the marginal of $\sum_\tau A[s,s^\prime;n,
n^\prime\vert\tau] P(\tau)$. The simplified form of the probability distributions (\ref{def2:Psn}) and (\ref{def2:Asn}) allows us to reduce our dynamic equation (\ref{eq2:diffusion}) to a system of ordinary differential equations (see \ref{section2:ODE} for details)
\begin{eqnarray}
 &&\hspace{-18mm} \frac{\mathrm d}{\mathrm d t}P_t(s,n)= \frac{1}{2}\left [1\!+\!s\tanh[\beta Jn+\beta\theta]\right
]P_t(-s,n)-\frac{1}{2}\left [1\!-\!s\tanh[\beta Jn+\beta\theta]\right]P_t(s,n)\nonumber\\
&&+pk\sum_{n^\prime=-k+1}^{k-1}A_t[s,1;n+1,n^{\prime}\vert1]~\frac{1}{2}[1-\tanh[\beta J(n^{\prime}\!+s)+\beta\theta]]\nonumber\\
&&+pk\sum_{n^\prime=-k+1}^{k-1}A_t[s,-1;n-1,n^{\prime}\vert1]~\frac{1}{2}[1+\tanh[\beta J(n^{\prime}\!+s)+\beta\theta]]\nonumber\\
&&-pk\sum_{n^\prime=-k+1}^{k-1}A_t[s,-1;n+1,n^{\prime}\vert1]~\frac{1}{2}[1+\tanh[\beta J(n^{\prime}\!+s)+\beta\theta]]\nonumber\\
&&-pk\sum_{n^\prime=-k+1}^{k-1}A_t[s,1;n-1,n^{\prime}\vert1]~\frac{1}{2}[1-\tanh[\beta J(n^{\prime}\!+s)+\beta\theta]].
\label{eq2:ODE}
\end{eqnarray}
Here $n\in\{-k,\ldots,k\}$, and $A_t[s,s^\prime ;n,n^\prime\vert1]=0$ for $n,n^\prime\notin\{-k+1,\ldots,k-1\}$, leading to four boundary equations.
The equations of the dynamical replica theory (\ref{eq2:WdilFt},\ref{eq2:Psn},\ref{eq2:Asn}) and (\ref{eq2:ODE}) are now cast into a form which allows us to solve them numerically.
\subsection{Numerical results}\label{section2:Numerics}
%

\begin{figure}[t]
\vspace*{0mm} \hspace*{-9mm} \setlength{\unitlength}{0.43mm}
\begin{picture}(350,210)

 \put(0,110){\includegraphics[height=100\unitlength,width=160\unitlength]{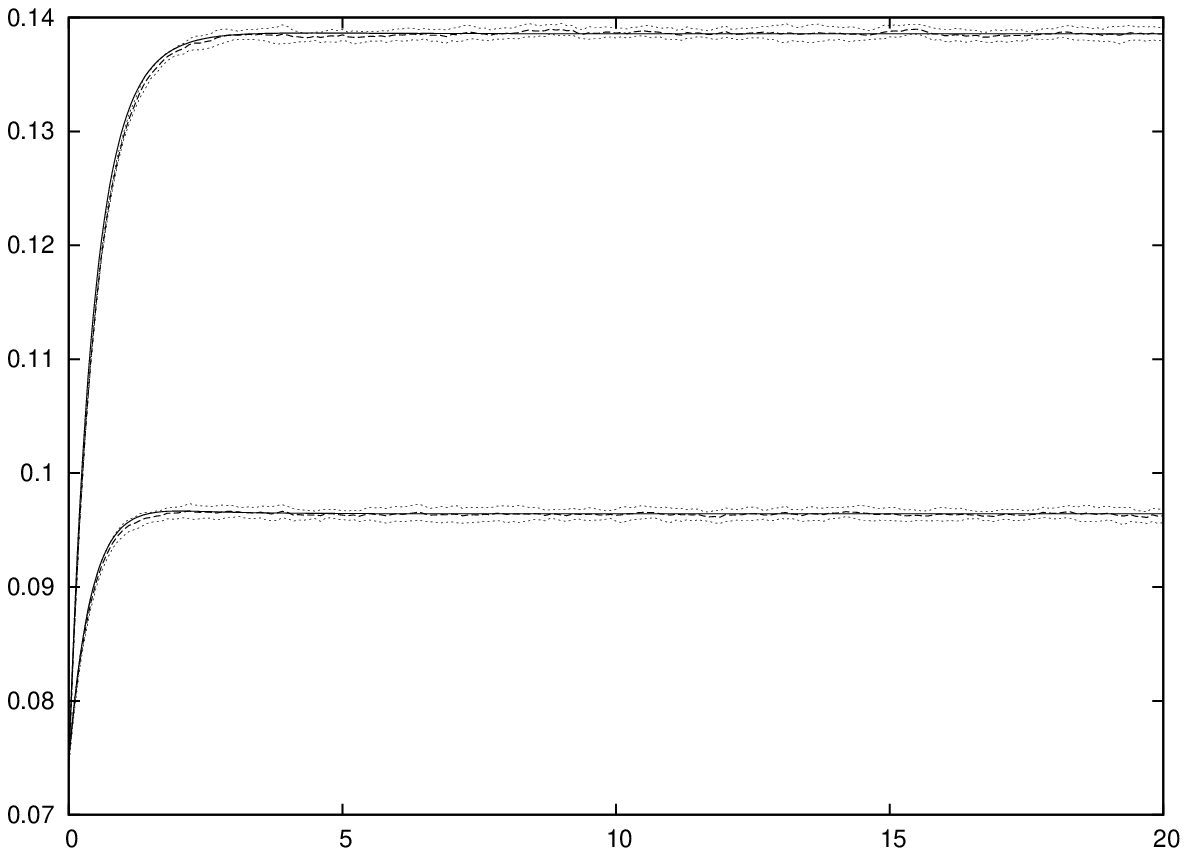}}
 \put(-1,65){\small\here{$m$}}
 \put(155,110){\includegraphics[height=100\unitlength,width=180\unitlength]{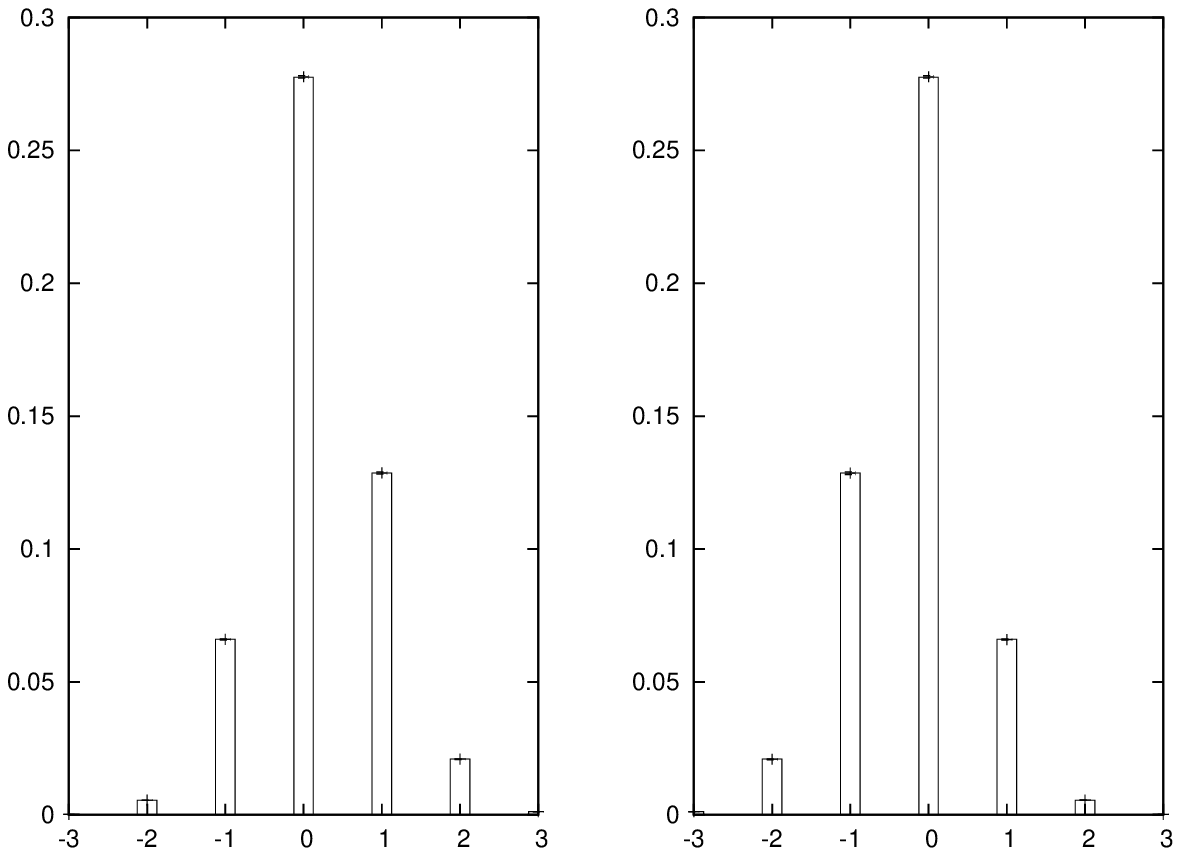}}
 \put(-1,165){\small\here{$\minus E$}}

 \put(0,0){\includegraphics[height=100\unitlength,width=160\unitlength]{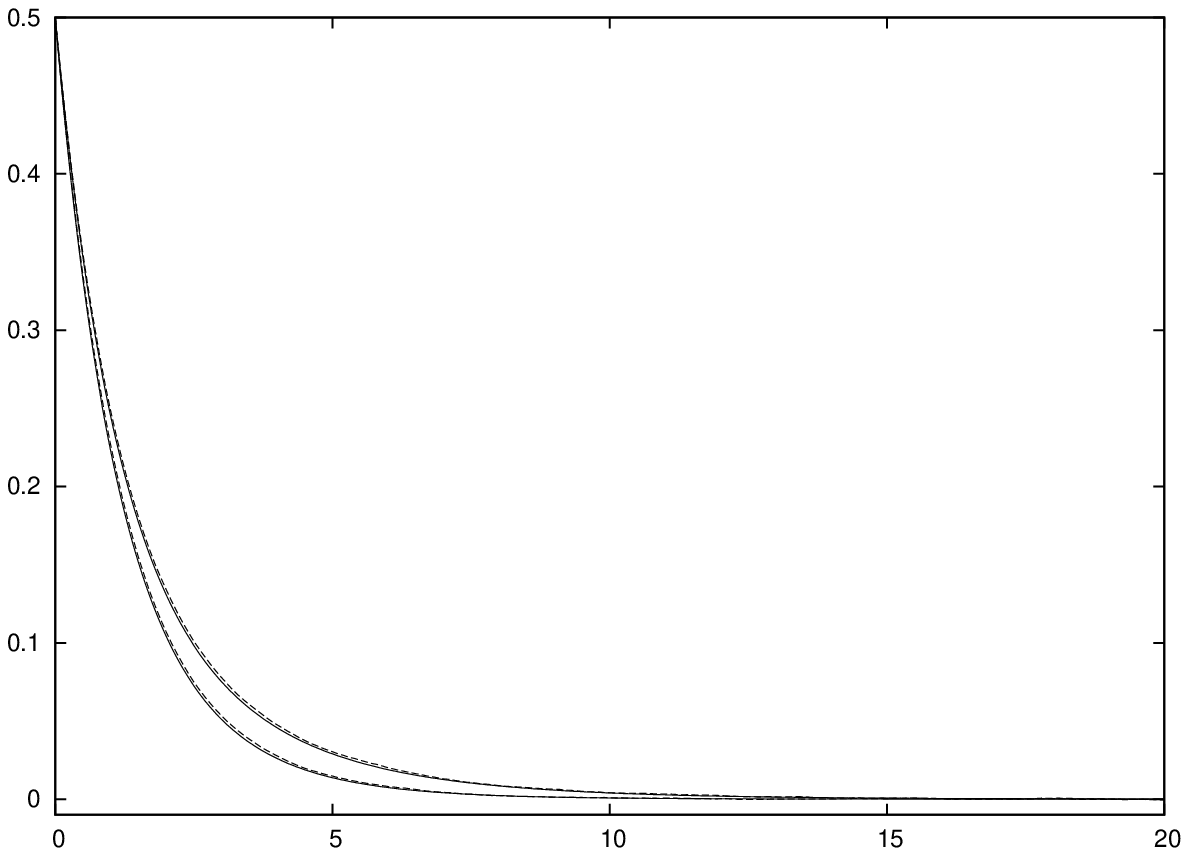}}
\put(155,0){\includegraphics[height=100\unitlength,width=180\unitlength]{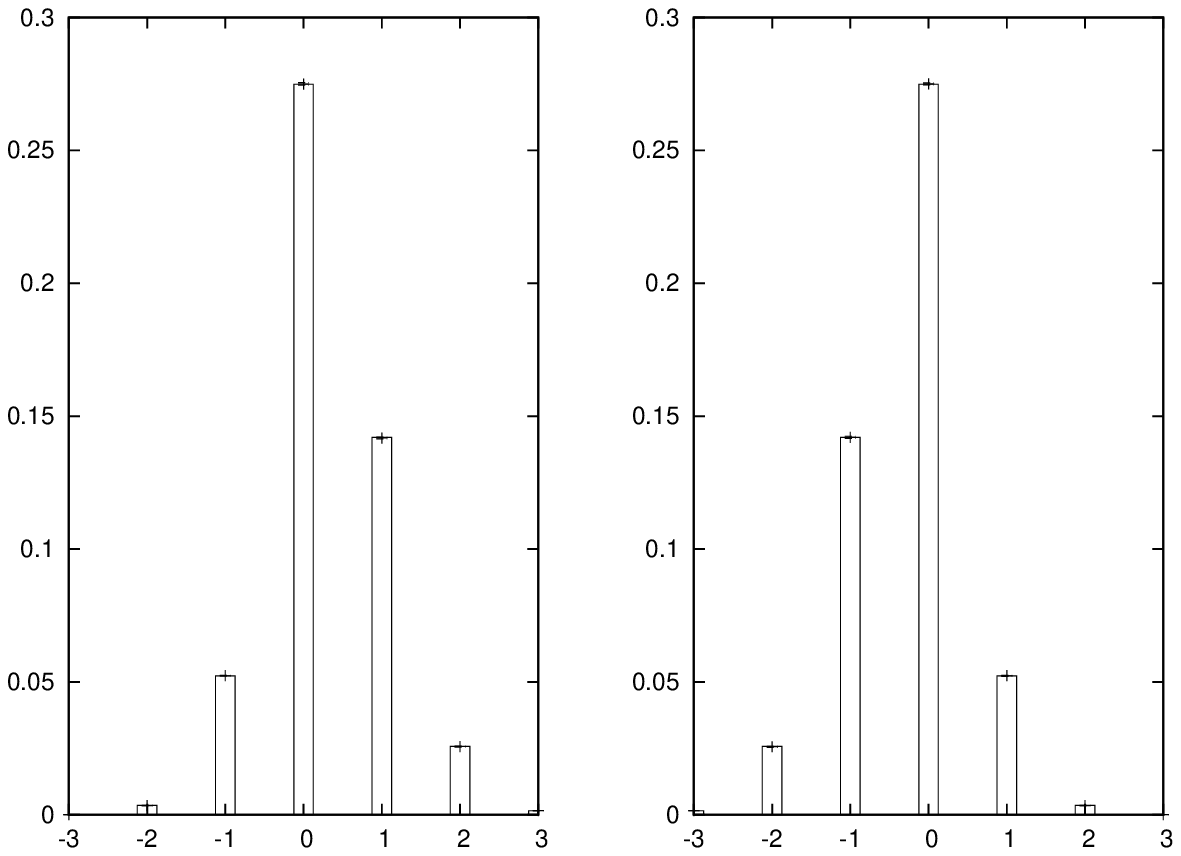}}
  \put(82,-7){\small{$t$}}   \put(203,-7){\small$n$} \put(294,-7){\small$n$}
\put(187,196){\here{\small $P(1,n)$}} \put(278,196){\here{\small $P(\!-\!1,n)$}}

\end{picture}
 \vspace*{0mm}
\caption{Left: evolution of the magnetization and energy per spin for $k=3$, $p=0.2$, $J=1$ and $\theta=0$. The temperatures are $T=3$ (bottom lines) and $T=2$ (top lines); the system is therefore in the paramagnetic phase since the Griffiths is found for $T\leq T_c(1)\approx 1.8205$. Time is measured in updates per spin. Solid lines represent results of the RS theory. Dashed and dotted lines denote the averages and averages $\pm$ standard deviation, respectively, as measured over $20$ MC simulations of systems with $N=10^6$ spins. For clarity we plot only the average MC magnetization. The size of the symbols is smaller than the error bars. Right: histograms (RS theory) of the two field distributions $P(\pm1,n)$ measured at $t=20$ compared to the corresponding MC results (markers with error bars). The top and bottom panels refer to the temperatures $T=3$ and $T=2$, respectively. } \label{fig:T3T2}
\end{figure}

\begin{figure}[t]
\vspace*{0mm} \hspace*{-9mm} \setlength{\unitlength}{0.43mm}
\begin{picture}(350,210)

 \put(0,110){\includegraphics[height=100\unitlength,width=160\unitlength]{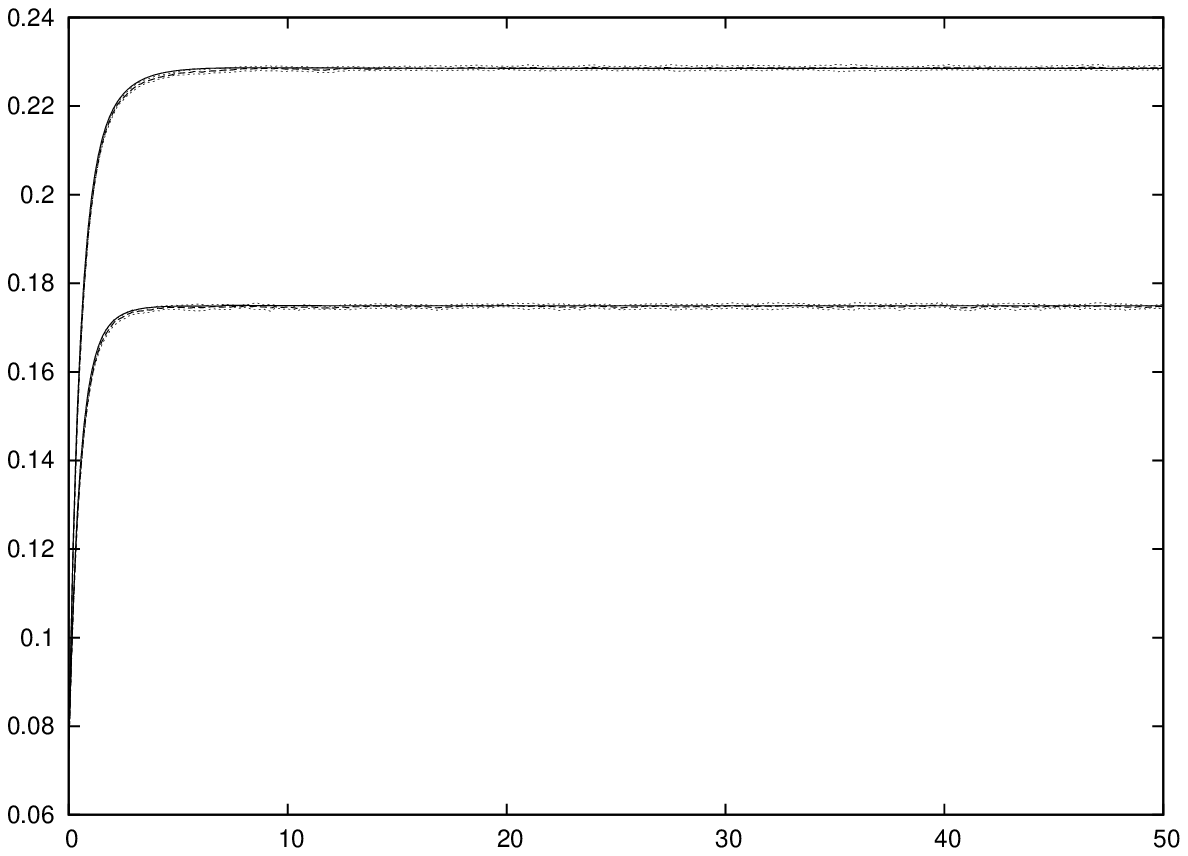}}
 \put(-1,65){\small\here{$m$}}
 \put(155,110){\includegraphics[height=100\unitlength,width=180\unitlength]{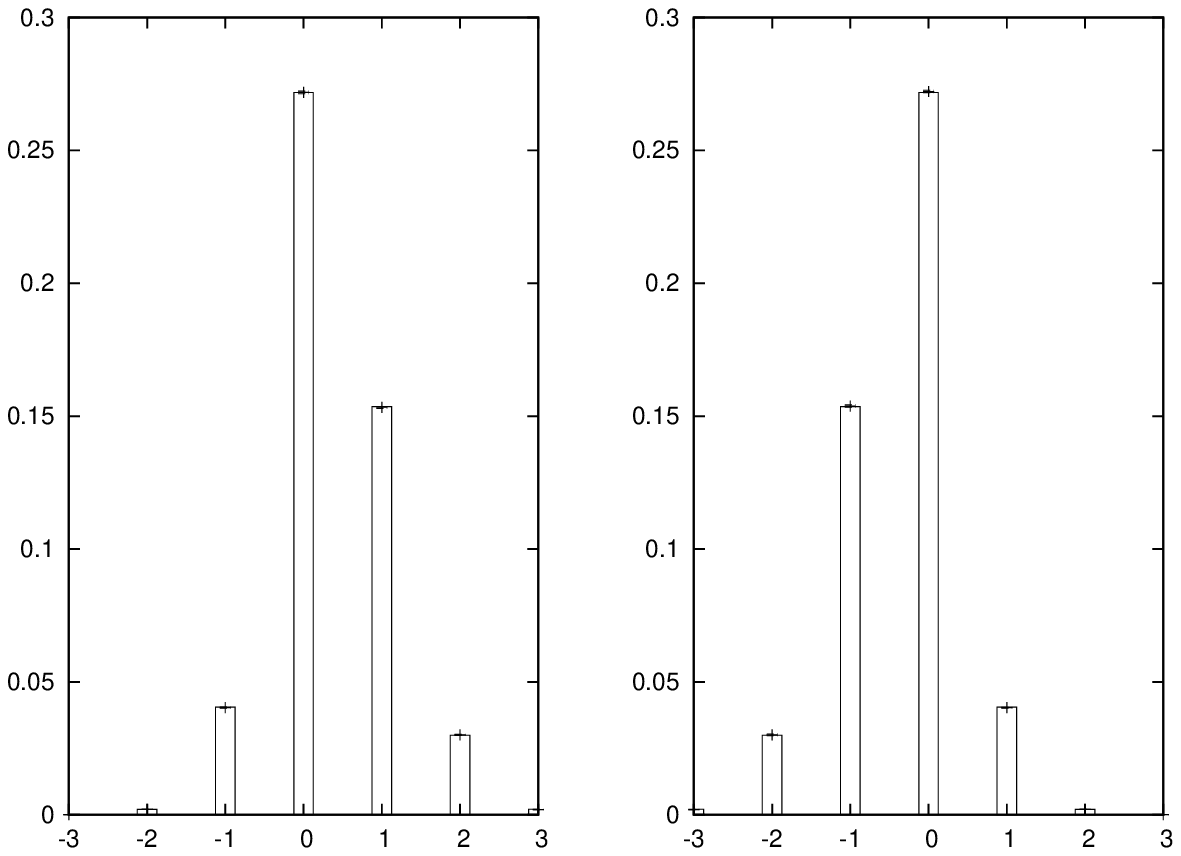}}
 \put(-1,165){\small\here{$\minus E$}}

 \put(0,0){\includegraphics[height=100\unitlength,width=160\unitlength]{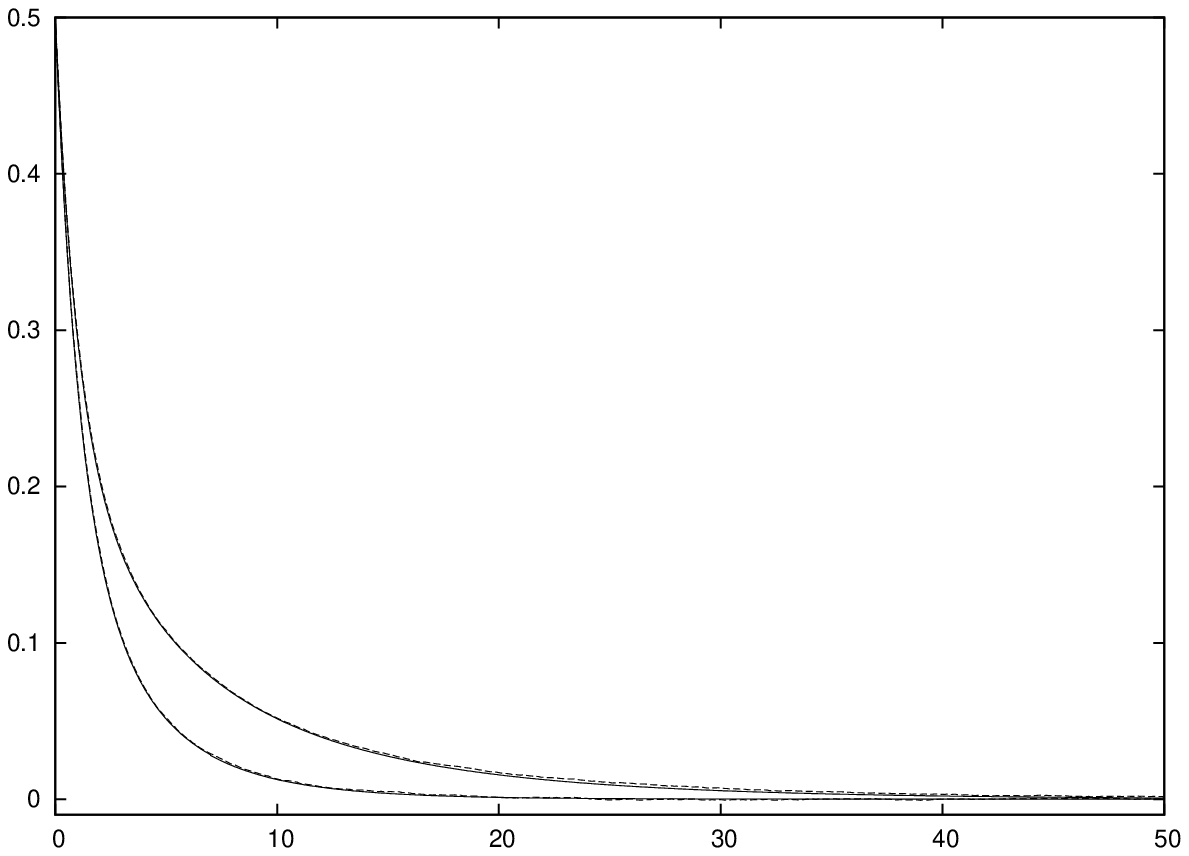}}
\put(155,0){\includegraphics[height=100\unitlength,width=180\unitlength]{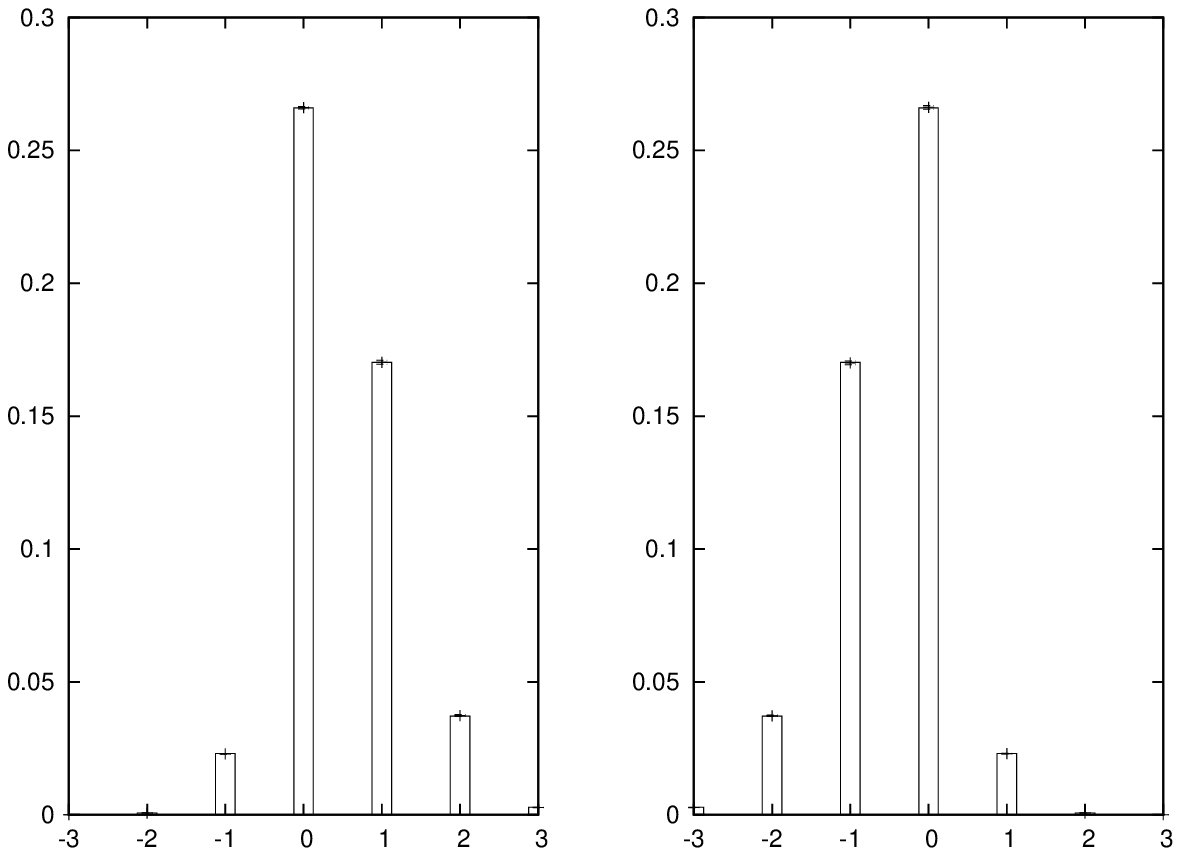}}
  \put(82,-7){\small{$t$}}   \put(203,-7){\small$n$} \put(294,-7){\small$n$}
\put(187,196){\here{\small $P(1,n)$}} \put(278,196){\here{\small $P(\minus1,n)$}}
\end{picture}
 \vspace*{0mm}
\caption{Left: evolution of the magnetization and energy per spin for $k=3$, $p=0.2$, $J=1$ and $\theta=0$. The temperatures are $T=1.5$ (bottom lines) and $T=1$ (top lines), so the system has entered the Griffiths phase.
Time is measured in updates per spin. Solid lines represent results of the RS theory. Dashed and dotted lines denote the averages and averages $\pm$ standard deviation, respectively, as measured over $20$ MC simulations of systems with $N=10^6$ spins. For clarity we plot only the average MC magnetization. The size of symbols is smaller than the error bars. Right: histograms (RS theory) of the two field distributions $P(\pm1,n)$ measured at $t=50$ compared to the corresponding MC results (markers with error bars). The top and bottom panels refer to the temperatures $T=1.5$ and $T=1$ respectively.} \label{fig:T15T1}
\end{figure}
\begin{figure}[t]
\vspace*{0mm} \hspace*{-9mm} \setlength{\unitlength}{0.43mm}
\begin{picture}(350,210)

 \put(0,110){\includegraphics[height=100\unitlength,width=160\unitlength]{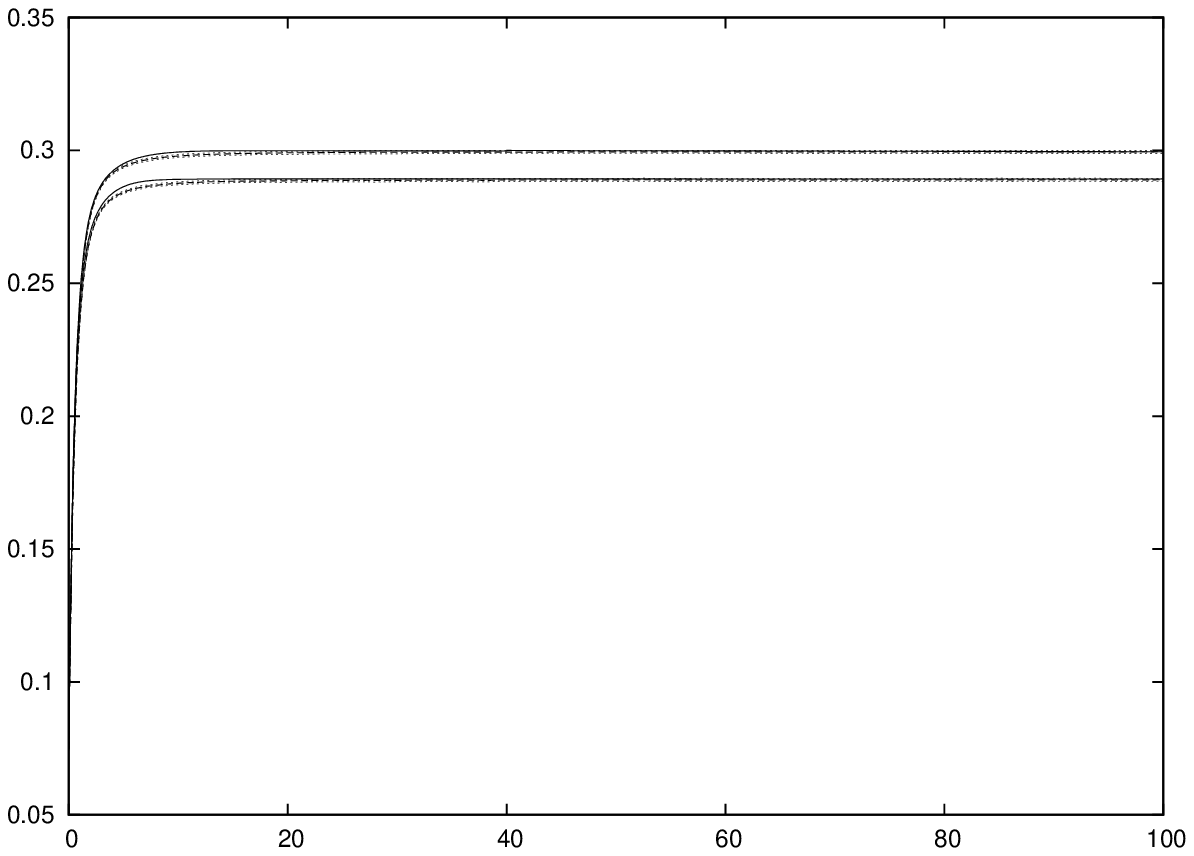}}
 \put(-1,65){\small\here{$m$}}
 \put(155,110){\includegraphics[height=100\unitlength,width=180\unitlength]{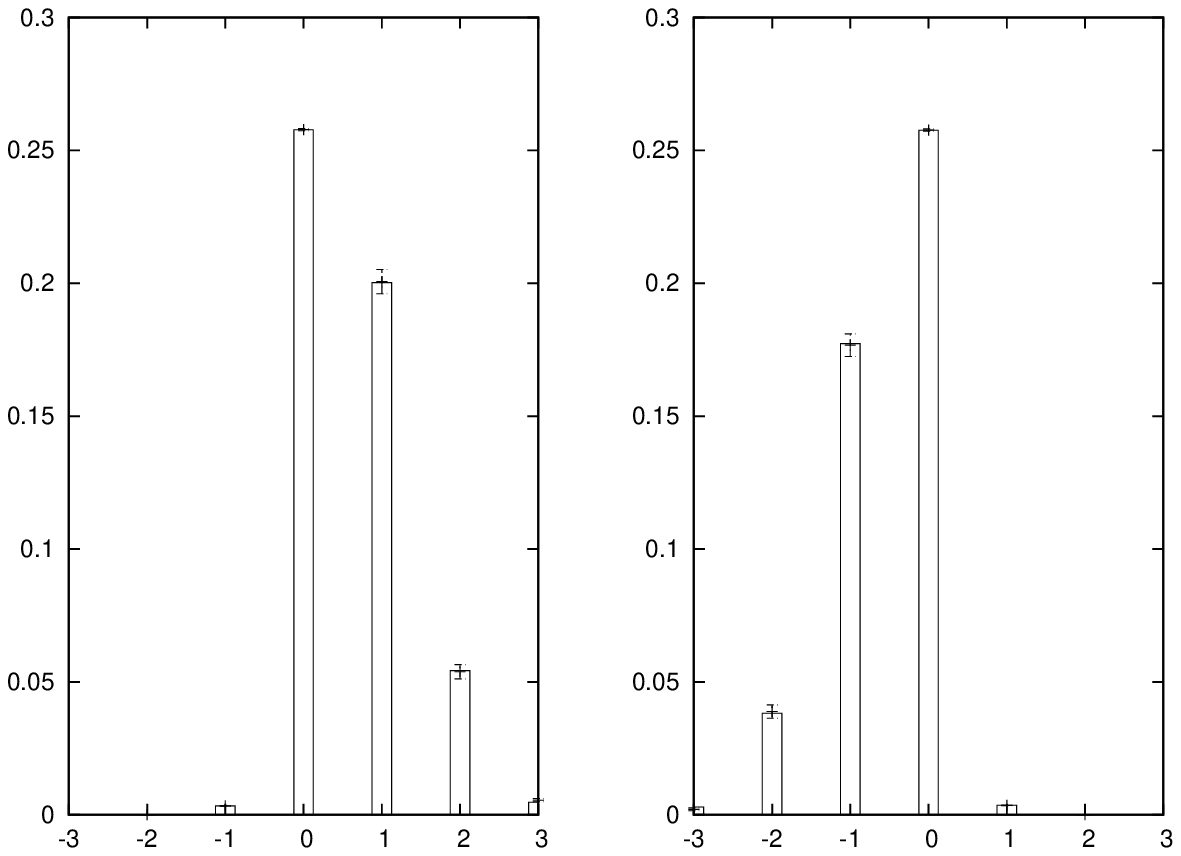}}
 \put(-1,165){\small\here{$\minus E$}}

 \put(0,0){\includegraphics[height=100\unitlength,width=160\unitlength]{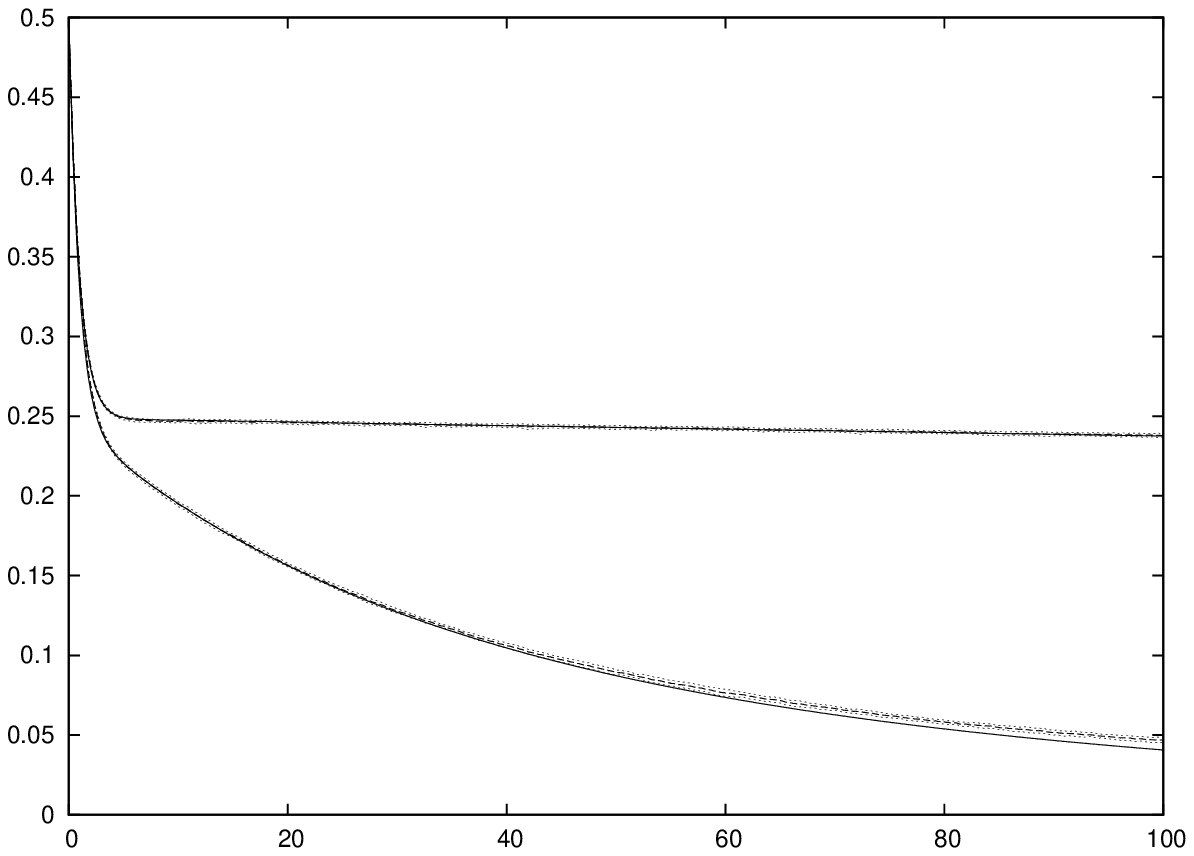}}
\put(155,0){\includegraphics[height=100\unitlength,width=180\unitlength]{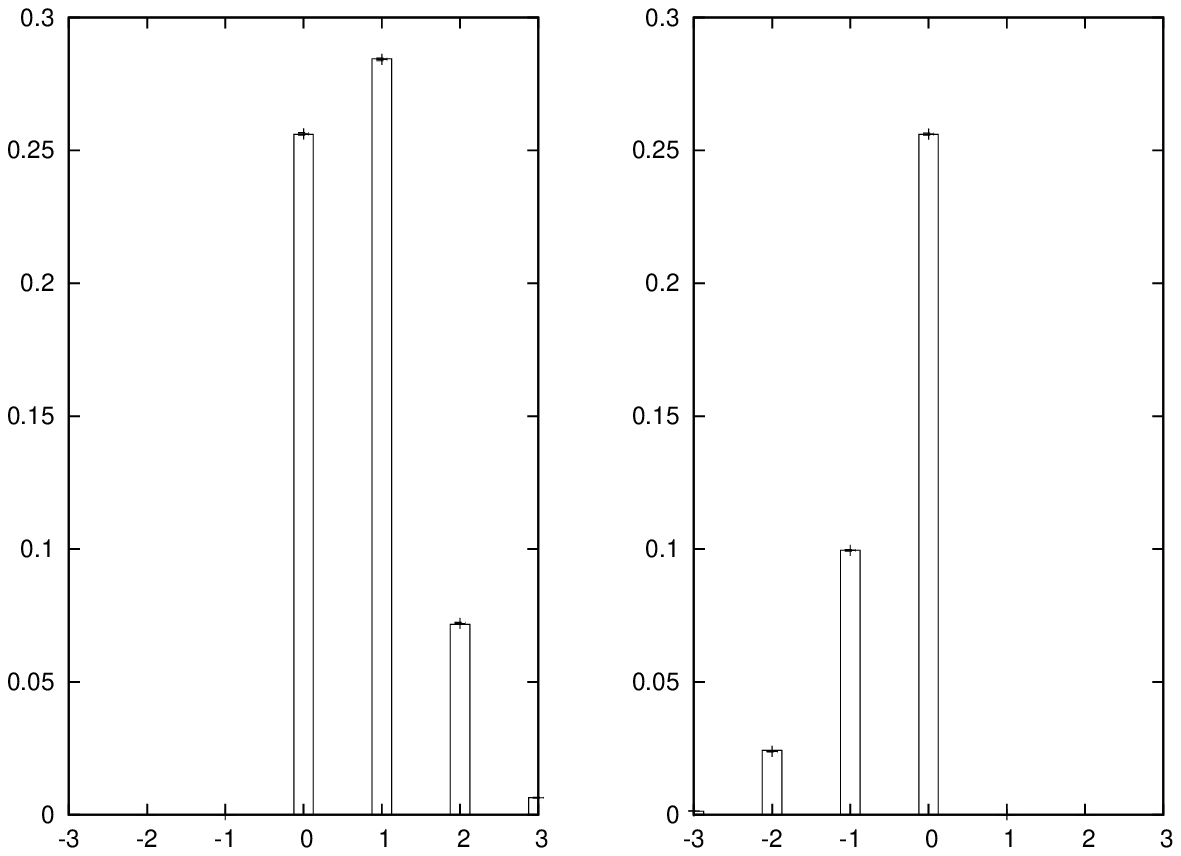}}
  \put(82,-7){\small{$t$}}   \put(203,-7){\small$n$} \put(294,-7){\small$n$}
\put(187,196){\here{\small $P(1,n)$}} \put(278,196){\here{\small $P(\minus1,n)$}}
\end{picture}
 \vspace*{0mm}
\caption{Left: evolution of the magnetization and energy per spin for $k=3$, $p=0.2$, $J=1$ and $\theta=0$. The temperatures are $T=0.5$ (bottom lines) and $T=0.25$ (top lines), so we have entered further into the Griffiths phase. Time is measured in updates per spin. Solid lines represent results of the RS theory. Dashed and dotted lines denote the averages and averages $\pm$ standard deviation, respectively, as measured over $20$ MC simulations of systems with $N=10^6$ spins. Right: histograms (RS theory) of the two field distributions $P(\pm1,n)$, measured at $t=100$, compared to the corresponding MC results (markers with error bars). The top and bottom panels refer to the temperatures $T=0.5$ and $T=0.25$ respectively.} \label{fig:T05T025}
\end{figure}
\begin{figure}[t]
\vspace*{5mm} \hspace*{35mm} \setlength{\unitlength}{0.58mm}
\begin{picture}(100,100)
\put(0,0){\epsfysize=105\unitlength\epsfbox{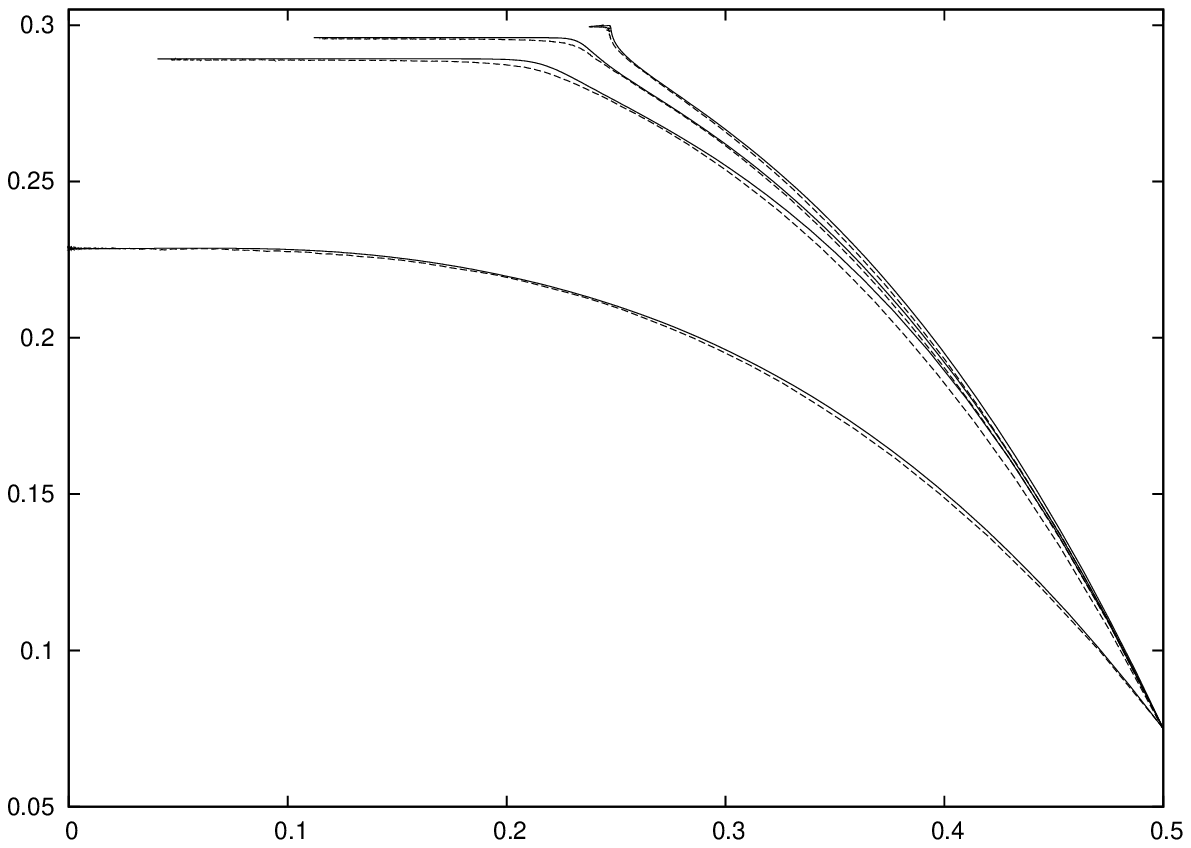}}
\put(80,-8){\here{\small $m$}}
\put(-6,55){\here{\small $-E$}}

\end{picture}
\vspace*{8mm} \caption{Evolution in time of the energy per spin $E$ and the magnetization $m$, now shown as trajectories in the $(m,E)$ plane,
for $k=3$, $p=0.2$, $J=1$, $\theta=0$ and temperatures $T=1, 0.5, 0.4, 0.25$ (from bottom to top), all of which correspond to the Griffiths phase.
Solid lines represent the predictions of the RS theory. Dashed lines denote average values measured over $20$ MC simulations of systems with $N=10^6$ spins each. The simulations were run for $100 N$ sequential spin updates, for all temperatures,
and the theoretical predictions calculated for the equivalent real-time duration $t\in[0,100]$. } \label{fig:EvsM}
\end{figure}


%
Here we use the analytic results of the previous section to study the dynamics in the Griffiths phase of the diluted Ising ferromagnet (\ref{def2:Ediluted}).
We solve the dynamical equation (\ref{eq2:ODE})  for the probability distribution $P_t(s,n)$ numerically, given the initial values (see \ref{section2:initial}) and given the boundary conditions of this equation, using Euler's forward iteration method. At each iteration step of this method we solve equations (\ref{eq2:WdilFt}) and (\ref{eq2:Psn}), using a population dynamics algorithm (see \ref{section2:pop} ), for the distribution $\tilde W$ and the function $d$. The result is then used to compute the probability distribution (\ref{eq2:Asn}) and to iterate the discrete version of the dynamic equation (\ref{eq2:ODE}) over the next time step $t\rightarrow t+\Delta t$.
In order to assess the quality of our dynamic theory we compare results of our numerical solutions of (\ref{eq2:ODE}) with the results of Monte Carlo (MC) simulations. In each simulation we generate a random regular graph of degree $k$ with $N$ vertices using the algorithm of Steger and Wormald \cite{StegerWormald}. We then remove each of the edges from this graph with probability $1-p$, so that on average only $p k N/2$ edges remain in the resulting diluted random graph. Finally, we perform MC simulations of the ferromagnetic Ising model defined on the diluted random graph (\ref{def2:Ediluted}) using conventional Glauber dynamics.

The evolution in time of the magnetization and the energy per spin, as obtained firstly in the numeric solution of theory and secondly
in the MC simulations, is depicted and compared in figures \ref{fig:T3T2}-\ref{fig:T05T025}. In addition we also compare in these figures the theoretical predictions for the histograms of fields $P_t(s,n)$ with the corresponding MC results, as measured in the final stage of each simulation. We observe that the theory correctly predicts both the trajectories of the macroscopic observables and the distributions of fields obtained in the MC simulations. Furthermore, one clearly notices the profound differences between the macroscopic dynamics of the model (\ref{def2:Ediluted}) in the paramagnetic phases (figure \ref{fig:T3T2}) versus the Griffiths phase  (figure \ref{fig:T05T025}).

The mesoscopic picture usually put forward to understand the dynamics of spin systems in the Griffiths phase \cite{Vojta} is that of local spin clusters that can be regarded as independent from (or only weakly dependent on)  the rest of the system. Each such cluster behaves as a finite (size $n$) local ferromagnet, with  its own 'local' ordering temperature $T_n$. A cluster of size $n$ is more likely to be found in the disordered $m_n=0$ state (where $m_n$ is its magnetization) above $T_n$, and in an ordered $m_n\neq0$ state for $T<T_n$.
At low temperatures the cluster is equally likely to be in one of its two ground states $\pm m_n$, which are related by the reversal $\sigma_i\rightarrow -\sigma_i$ of all spins in the cluster. In order to go from $m_n$ to $-m_n$ the cluster has to overcome an energy barrier $E_n$. The microscopic time $\tau_n$ required for this operation to occur is given by the Arrhenius form $\tau_n\sim\exp[-E_n/T]$. The collective behavior of these clusters is thought to be responsible for the slowing down of the dynamics in the Griffiths phase \cite{Vojta}.

The above picture indeed allows us to interpret the results of the present study. Our numerical results (figures \ref{fig:T3T2}-\ref{fig:T05T025})
refer to regular random graphs of degree $k=3$, with dilution strength $p=\frac{1}{5}$, which is below the percolation threshold $p_c=\frac{1}{2}$ for this graph. The simulated system therefore consists of independent clusters of finite size, and the density $W_n(p)$ of large clusters decays exponentially according to (\ref{eq2:clusterDistr}). The Griffiths phase of the model (\ref{def2:Ediluted}) (for $k=3$, $J=1$ and $p=\frac{1}{5}$) is the range of temperatures $0<T<T_c(1)$, where $T_c(1)=1.820478(6)$ is the critical temperature of the corresponding 'clean' undiluted system. Above
$T_c(1)$ all clusters are paramagnetic, and the magnetization and the energy both relax quickly to their equilibrium values $m=0$ and (\ref{eq2:Eeq}),  respectively (see figure \ref{fig:T3T2}). The distribution of fields $P(s,n)$ (see figure \ref{fig:T3T2}) is symmetric, i.e. $P(s,n)=P(-s,-n)$, as it should be in equilibrium when $\theta=0$ in (\ref{def2:Ediluted}). In the Griffiths phase, in contrast, both paramagnetic and ferromagnetic clusters are present. For short times the paramagnetic and ferromagnetic clusters  evolve to the $m_n=0$ and $m_n\neq0$ states, respectively. At intermediate times the magnetizations of paramagnetic clusters simply fluctuate around $m_n=0$, whereas the ferromagnetic clusters will 'flip' $m_n\rightarrow-m_n$ as frequently as the relaxation time of the cluster $\tau_n$ allows. Larger clusters require more time to 'flip' due to their energy barriers being proportional to their sizes. Furthermore, for lower temperatures the ferromagnetic clusters tend to stay longer in each of their two ground states $\pm m_n$. Eventually the whole system ends up in the zero global magnetization state. However, how quickly this would happen depends on the control parameters of the system. In the high temperature region of the Griffiths phase the ferromagnetic clusters would 'flip' frequently, and although the relaxation time of the order parameters decreases at lower temperatures, it is still relatively quick; see figure \ref{fig:T15T1}. We also observe in this latter figure that the energy attains its equilibrium value, given by (\ref{eq2:Eeq}), much earlier than the magnetization. This marks the onset of the main stage of the dynamics, where only the flips $m_n\leftrightarrow-m_n$ of the ferromagnetic clusters are relevant. As we decrease the temperature further the dynamics becomes very slow, see figure \ref{fig:T05T025}. Here the energy has attained its equilibrium value, but the magnetization has not.  The number of ferromagnetic clusters has increased, and so have the relaxation times $\tau_n$ of those clusters which were already ferromagnetic at higher temperatures. Furthermore, at $T=0.25$ we observe that in the MC simulation the equilibration times diverge with the system size $N$ (at $T=0.5$, in contrast, the system can be still equilibrated on timescales significantly less than the system size). This suggests the existence of another critical temperature $T_\star$, which for the parameters of the system in this study would be located somewhere in the interval $0.5<T_\star<0.25$, that separates the Griffiths phase into two further distinct regions of relatively slow and relatively fast dynamics, respectively. A possible mechanism behind this further (dynamic) transition would be that the number of clusters which are ferromagnetic becomes extensive, combined with diverging cluster relaxation times. Interestingly, the flow in the energy versus magnetization plane (see figure \ref{fig:EvsM}) for the temperature $T=0.25$ is distinct from that observed at higher temperatures, in terms of an apparently discontinuous direction  change, and this is observed in both theory and simulation. In contrast, in the paramagnetic and high temperature Griffiths regions of this model the  trajectories in the $(m,E)$ plane are smooth. It is not yet clear to what extent the temperature at which the distinct direction of $(m,E)$ flow sets in is related to the suggested dynamic transition temperature $T_\star$ of diverging relaxation times.

\section{Summary and conclusions}\label{section2:Discussion}
In this paper we built on a recent study \cite{MozeikaCoolen} in which a dynamical replica theory (DRT) was developed to solve the (sequential) stochastic dynamics of finitely connected Ising spin systems with random bonds. Here we generalized this theory to  include systems on random graphs defined by {\em arbitrary} vertex degree distributions (as apposed to the Poissonnian ones of \cite{MozeikaCoolen}). We have used the exact dynamical  equation for the joint spin-field probability distribution, that was derived in \cite{MozeikaCoolen}, as a starting point. We closed this equation following the standard assumptions of DRT. The resulting macroscopic theory takes the form  of a nonlinear diffusion equation coupled to a functional saddle-point problem, where the latter involves replica density order parameters that are to be solved at each instance of time. We showed that the results of equilibrium statistical mechanics \cite{FCNN} can be recovered within our dynamic theory,  and that the equilibrium solution of the model is a stationary point of our macroscopic equations. The saddle-point equations resulting from making a replica-symmetric ansatz can be solved numerically by a population dynamics algorithm \cite{BetheSG}. The results in \cite{MozeikaCoolen} for random graphs with Poissonian degree distributions are easily recovered from our generalized equations.

We have applied our theory to the dynamics of the diluted Ising ferromagnet in the Griffiths phase. This model is an Ising ferromagnet defined on a random regular graph from which edges are removed randomly, with some probability $1-p$. The local fields in this model take integer values, which simplified our dynamic theory to a system of ordinary differential equations for the joint probability distribution of Ising spins and integer fields. The functional order parameter of the saddle-point problem is a distribution over real-valued $2\times2$  matrices. We have solved our dynamic equations numerically for random regular graphs of degree $k=3$ with dilution $p=\frac{1}{5}$, and calculated the evolution in time of the magnetization and the energy per spin in both the paramagnetic and the Griffiths phases of this model. Dynamic Griffiths effects are clearly present in the Griffiths phase. The magnetization equilibrates much slower than the energy, and this discrepancy becomes even more severe in the low temperature region of the Griffiths phase.  In contrast to the paramagnetic phase and higher temperature region of the Griffiths phase, the energy per spin appears to be no longer a smooth function of the magnetization in the low temperature region of the Griffiths phase. The equilibration times of the MC simulation, the results of which are in good agreement with the numeric solutions of our theory, diverge with the system size in the low temperature region of the Griffiths phase.

The predictions of the dynamical theory presented in this paper for the diluted Ising ferromagnet in its Griffiths phase are remarkably accurate. To us this is not entirely surprising, for at least two reasons. First, the dynamic replica theory and its variations have in the past already proven to be very accurate for ferromagnetic models on regular \cite{ApprSch,DRTfc} and Poissonian \cite{MozeikaCoolen} random graphs. Second, the extended version of DRT considered here describes the evolution of the joint spin-field probability distribution. In equilibrium, the cluster expansion of the magnetization derived in \cite{Harris} can be recovered within the cavity approach \cite{Laumann}, which is equivalent to the replica method. This suggests that the joint spin-field probability indeed contains the relevant information about the clusters which is responsible for the slow dynamics in the Griffiths phase.

\section*{Acknowledgements}
We are very grateful to Nick Wormald for his helpful suggestions on the numerical generation of random graphs.

\appendix
\section{Averaging over disorder}\label{section2:average}
In this section we compute averages over disorder $\{c_{ij}J_{ij}\}$ in the equation (\ref{eq2:All}). First, we exploit the $i\leftrightarrow j$ symmetry of the interactions $c_{ij}J_{ij}$ to write the disorder dependent term of this equation in form more convenient for further manipulations. Second, we write the Kronecker delta in the definition of the connectivity $\{c_{ij}\}$ disorder (\ref{eq2:P(C)}) in its integral representation (\ref{def2:Kdelta}). This gives us
\begin{eqnarray}
&&\hspace{-25mm}\bra\ldots\ket_{\{c_{ij}J_{ij}\}}
=\rme^{-\rmi\theta\sum_{\alpha,i}\hat{h}_i^{\alpha}}\left\langle
c_{\ell\ell^\prime}\delta [ h\! -\!H_{\ell^\prime}^1\!+\!2J_{\ell\ell^\prime}\tilde s]\rme^{-\rmi\sum_{i< j}
c_{ij}J_{ij}\sum_{\alpha}\big\{\hat{h}_i^{\alpha}
\sigma_j^{\alpha}+\hat{h}_j^{\alpha}
\sigma_i^{\alpha}\big\}}\right\rangle_{\{c_{ij}J_{ij}\}}\nonumber\\
&&\hspace{-8mm}=\frac{1}{\Z}\rme^{-\rmi\theta\sum_{\alpha , i}\hat{h}_i^{\alpha}}\sum_{\cv}
\prod_{i<j}\left[\frac{c}{N}\delta_{c_{ij},1}+
(1\!-\!\frac{c}{N})\delta_{c_{ij},0}\right]\nonumber\\
&&\hspace{-5mm}\times\prod_i\int_{-\pi}^{\pi}\frac{\mathrm d \hat k_i}{2\pi}~\rme^{\rmi\hat k_i(k_i-\sum_{j\neq i} c_{ij})}\nonumber\\
&&\hspace{-5mm}\times \left\langle c_{\ell\ell^\prime}\delta [ h
-H_{\ell^\prime}^1+2J_{\ell\ell^\prime}\tilde s]~\rme^{-\rmi\sum_{i< j}
c_{ij}J_{ij}\sum_{\alpha}\big\{\hat{h}_i^{\alpha}
\sigma_j^{\alpha}+\hat{h}_j^{\alpha}
\sigma_i^{\alpha}\big\}}\right\rangle_{\{J_{ij}\}}.
\end{eqnarray}
 Taking the average over connectivity disorder $\{c_{ij}\}$ leads us to
\begin{eqnarray}
&&\hspace{-25mm}\langle\ldots\rangle_{\{c_{ij}J_{ij}\}}=\frac{1}{\Z}\rme^{-\rmi\theta\sum_{\alpha , i}\hat{h}_i^{\alpha}}\!\int_{-\pi}^{\pi}\prod_i\left[\frac{\mathrm d \hat k_i}{2\pi}~\rme^{\rmi\hat k_i k_i}\right]\sum_{\cv}
\prod_{i<j}\left[\frac{c}{N}\delta_{c_{ij},1}+
(1-\frac{c}{N})\delta_{c_{ij},0}\right]\nonumber\\
&&\hspace{-4mm}\times\! \left\langle\! c_{\ell\ell^\prime}\delta [ h\!
-\!H_{\ell^\prime}^1\!+\!2J_{\ell\ell^\prime}\tilde s]\rme^{-\rmi\sum_{i< j}
c_{ij}\left[J_{ij}\!\sum_{\alpha}\big\{\hat{h}_i^{\alpha}
\sigma_j^{\alpha}+\hat{h}_j^{\alpha}
\sigma_i^{\alpha}\big\}-\rmi\lbrace\hat k_i+ \hat k_j\rbrace\right]}\!\right\rangle_{\!\{J_{ij}\}}\nonumber\\
&&\hspace{-13mm}=\frac{1}{\Z}\rme^{-\rmi\theta\sum_{\alpha , i}\hat{h}_i^{\alpha}}\int_{-\pi}^{\pi}\prod_i\left[\frac{\mathrm d \hat k_i}{2\pi}~\rme^{\rmi\hat k_i k_i}\right]\nonumber\\
&&\hspace{-10mm}\times\frac{c}{N}\int \mathrm d J~ P(J)~\delta [ h
-H_{\ell^\prime}^1+2J\tilde s]~\rme^{-\rmi J \sum_{\alpha}\big\{\hat{h}_{\ell}^{\alpha}
\sigma_{\ell^\prime}^{\alpha}+\hat{h}_{\ell^\prime}^{\alpha}
\sigma_{\ell}^{\alpha}\big\}-\rmi\lbrace\hat k_{\ell}+ \hat k_{\ell^\prime}\rbrace}\nonumber\\
&&\hspace{-10mm}\times
\prod_{i<j}\left\lbrace\frac{c}{N}\int \mathrm d J~ P(J)~\rme^{-\rmi
J\sum_{\alpha}\big\{\hat{h}_i^{\alpha}
\sigma_j^{\alpha}+\hat{h}_j^{\alpha}
\sigma_i^{\alpha}\big\}-\rmi\lbrace\hat k_i+ \hat k_j\rbrace}+(1\!-\!\frac{c}{N})\right\rbrace\label{eq2:av2}
\end{eqnarray}
where in the last line of above expression $i\neq\ell$ and $j\neq\ell^\prime$. Finally, upon re-exponentiating (\ref{eq2:av2}) we obtain our desired result for the disorder average in (\ref{eq2:All}):
\begin{eqnarray}
&&\hspace{-25mm}\langle\ldots\rangle_{\{c_{ij}J_{ij}\}}=~\frac{1}{\Z}\rme^{-\rmi\theta\sum_{\alpha , i}\hat{h}_i^{\alpha}}\int_{-\pi}^{\pi}\!\prod_i\left[\frac{\mathrm d \hat k_i}{2\pi}~\rme^{\rmi\hat k_i k_i}\right]\nonumber\\
&&\hspace{-20mm}\times\frac{c}{N}\int\! \mathrm d J~ P(J)~\delta [ h
-H_{\ell^\prime}^1+2J\tilde s]~\rme^{-\rmi J \sum_{\alpha}\big\{\hat{h}_{\ell}^{\alpha}
\sigma_{\ell^\prime}^{\alpha}+\hat{h}_{\ell^\prime}^{\alpha}
\sigma_{\ell}^{\alpha}\big\}-\rmi\lbrace\hat k_{\ell}+ \hat k_{\ell^\prime}\rbrace}\nonumber\\
&&\hspace{-20mm}\times
\exp\left[\frac{c}{2N}\sum_{i,j}\left\lbrace\int\! \mathrm d J P(J)\rme^{-\rmi
J\sum_{\alpha}\big\{\hat{h}_i^{\alpha}
\sigma_j^{\alpha}+\hat{h}_j^{\alpha}
\sigma_i^{\alpha}\big\}-\rmi\lbrace\hat k_i+ \hat k_j\rbrace}\!-\!1\right\rbrace+O(N^{0})\right].
\end{eqnarray}

\section{Computation of the kernel $A[\ldots]$}\label{section2:A}
In this appendix we give details of the calculation which leads to the path integral (\ref{eq2:Aintegral}). We insert our result for the term  $\left\langle c_{\ell\ell^\prime}A_{\ell\ell^\prime}[\ldots]\right\rangle_{\{c_{ij}J_{ij}\}}$  (\ref{eq2:All}) into the sum (\ref{def2:A}), which gives
\begin{eqnarray}
&&\hspace{-25mm}A[s,s^\prime;h,h^\prime;\tilde s]=\frac{1}{
  \Z~N^2}\sum_{\ell\neq\ell^\prime}\int\prod_{\tau h \alpha}\frac{\mathrm
d\hat{D}_{\alpha}(\tau,h
)}{2\pi/N}\int\!\! \prod_{\sigmav,
\hat{\hv},\hat k}\Big[\frac{\mathrm d \hat{P}(\sigmav , \hat{\hv},\hat k)\mathrm d
P (\sigmav , \hat{\hv},\hat k)}{2\pi/N}\Big]
\nonumber\\
&&\times\exp \left[\rmi
N\!\sum_{\tau ,h ,\alpha}\hat{D}_{\alpha}(\tau,h
)D(\tau ,h
)+\rmi N\!\sum_{\sigmav}\!\int\! \mathrm d \hat{\hv}\mathrm d \hat k\hat{P}(\sigmav ,
\hat{\hv},\hat k)P (\sigmav , \hat{\hv},\hat k)
\nonumber\right.\\
&&
\left.\hspace*{10mm}+\frac{1}{2}c N
\sum_{\sigmav,\sigmav^\prime}\int\! \mathrm d \hat{\hv}\mathrm d
\hat{\hv}^\prime \mathrm d \hat k \mathrm d \hat{k}^\prime P (\sigmav , \hat{\hv},\hat k)P (\sigmav^\prime\! ,
\hat{\hv}^\prime\!,\hat{k}^\prime)\times\ldots
\nonumber\right.\\
&&\left.
\hspace*{30mm}
\ldots\times\Big\langle ~\rme^{-\rmi J[\hat{\hv} .\sigmav^\prime+\hat{\hv}^{\prime} .\sigmav]-\rmi[\hat k+\hat{k}^\prime]}-1\Big\rangle_{\!J}+O(N^{0})\room
\right]
\nonumber\\
&&
\times\sum_{\sigmav^1}\ldots\sum_{\sigmav^n}\int\!\prod_{i}\Big\{\frac{
\mathrm d \Hv_i\mathrm d
\hat{\hv}_i}{2\pi}\Big\}\int_{-\pi}^{\pi}\prod_i\Big[\frac{\mathrm d \hat k_i}{2\pi}~\rme^{\rmi\hat k_i k_i}\Big]\nonumber\\
&&\times\exp\Big[\rmi\sum_{i}\hat{\hv}_i
.\left\{\Hv_i-\thetav\right\}\Big]\nonumber\\
&&
\times\exp\Big[ - \rmi\sum_{\tau ,h,\alpha}
\hat{D}_{\alpha}(\tau ,h
)\sum_i\delta_{\tau ,\sigma_i^{\alpha}}\delta\left[h
-H_i^{\alpha}\right]-\rmi\sum_i\hat{P}(\sigmav_i , \hat{\hv}_i, \hat k_i)\Big]
\nonumber\\
&&
\times\delta_{s^\prime\!,\sigma_\ell^1}\delta_{s,\sigma_{\ell^\prime}^1}\delta
[h^\prime\! -H_\ell^1]
 \Big\langle\delta [ h -H_{\ell^\prime}^1+2J\tilde
s]~\rme^{-\rmi J[\hat{\hv}_\ell .\sigmav_{\ell^\prime}+
\hat{\hv}_{\ell^\prime} .\sigmav_\ell]}\Big\rangle_{\!J}
\nonumber\\
&&\times \rme^{-\rmi[\hat k_\ell+ \hat k_{\ell^\prime}]}.
\label{eq2:comp-of-A}
\end{eqnarray}
Next we rescale the conjugate integration variables according to $\hat D_\alpha\rightarrow\hat D_\alpha\Delta h$, and define the function
\begin{eqnarray}
&&\hspace{-25mm}M [\Hv_i,\hat{\hv}_i,\sigmav_i\vert k_i,\hat k_i,\theta]=\frac{1}{2\pi}~\rme^{\rmi\hat{\hv}_i
.\left\{\Hv_i-\thetav\right\}}\nonumber\\
&&\hspace{12mm}\times\rme^{- \rmi\sum_{\tau ,h,\alpha}\Delta h
\hat{D}_{\alpha}(\tau ,h
)~\delta_{\tau ,\sigma_i^{\alpha}}\delta\left[h
-H_i^{\alpha}\right]+\rmi\hat k_i k_i-\rmi\hat{P}(\sigmav_i , \hat{\hv}_i, \hat k_i)}
\end{eqnarray}
insertion of which into (\ref{eq2:comp-of-A}), followed by further manipulations, leads us to
\begin{eqnarray}
&&\hspace{-25mm}A[s,s^\prime;h,h^\prime;\tilde s]
=\frac{1}{
  \Z}\Big[\frac{1}{2\pi}\Big]^N\int\{\mathrm d P \mathrm d\hat P \mathrm d\hat
D\}\nonumber\\
&&\hspace{-5mm}\times\exp \left[\rmi
N\sum_{\tau ,h ,\alpha}\!\Delta h\hat{D}_{\alpha}(\tau ,h
)D(\tau ,h
)+\rmi N\sum_{\sigmav}\!\int\! \mathrm d \hat{\hv}\mathrm d \hat k\hat{P}(\sigmav ,
\hat{\hv},\hat k)P (\sigmav , \hat{\hv},\hat k)\nonumber
\right.\\
&&
\left.
+\frac{1}{2}c N
\sum_{\sigmav,\sigmav^\prime}\int \mathrm d \hat{\hv}\mathrm d
\hat{\hv}^\prime \mathrm d \hat k \mathrm d \hat{k}^\prime P (\sigmav , \hat{\hv},\hat k)P (\sigmav^\prime ,
\hat{\hv}^\prime,\hat{k}^\prime)\times\ldots
\right.\nonumber\\
&&
\hspace*{40mm}
\left.
\ldots\times\Big\langle ~\rme^{-\rmi J[\hat{\hv} .\sigmav^\prime+\hat{\hv}^{\prime} .\sigmav]-\rmi[\hat k+\hat{k}^\prime]}-1\Big\rangle_{\!J}
\right.\nonumber\\
&&
\left.+\sum_i\log\sum_{\sigmav_{\!i}}\int\mathrm d \Hv_i\mathrm d
\hat{\hv}_i\int_{-\pi}^{\pi}\mathrm d \hat k_i~ M [\Hv_i,\hat{\hv}_i,\sigmav_i\vert k_i,\hat k_i,\theta]+O(N^{0})\room\right]
\nonumber
\\
&&\times\frac{1}{N^2}\sum_{\ell\neq\ell^\prime}\Bigg\{\sum_{\sigmav_{\!\ell}}\int
\mathrm d \Hv_\ell\mathrm d
\hat{\hv}_\ell\int_{-\pi}^{\pi}\mathrm d \hat k_\ell M [\Hv_\ell,\hat{\hv}_\ell,\sigmav_\ell\vert k_\ell,\hat k_\ell,\theta]\nonumber
\\
&&\times\sum_{\sigmav_{\!\ell^\prime}}\int
\mathrm d \Hv_{\ell^\prime}\mathrm d
\hat{\hv}_{\ell^\prime}\int_{-\pi}^{\pi}\mathrm d \hat k_{\ell^\prime} M [\Hv_{\ell^\prime},\hat{\hv}_{\ell^\prime},\sigmav_{\ell^\prime}\vert k_{\ell^\prime},\hat k_{\ell^\prime},\theta]\nonumber\\
&&\times\delta_{s^\prime,\sigma_\ell^1}\delta_{s,\sigma_{\ell^\prime}^1}\delta
[h^\prime -H_\ell^1]
 \left\langle\delta [ h -H_{\ell^\prime}^1+2J\tilde
s]~\rme^{-\rmi J[\hat{\hv}_\ell .\sigmav_{\ell^\prime}+
\hat{\hv}_{\ell^\prime} .\sigmav_\ell]}\right\rangle_J\nonumber\\
&&\times
 \rme^{-\rmi[\hat k_\ell+ \hat k_{\ell^\prime}]}\nonumber\\
&&\times\Big[\sum_{\sigmav_{\!\ell}}\int\!
\mathrm d \Hv_\ell\mathrm d
\hat{\hv}_\ell\int_{-\pi}^{\pi}\mathrm d \hat k_\ell M [\Hv_\ell,\hat{\hv}_\ell,\sigmav_\ell\vert k_\ell,\hat k_\ell,\theta]\nonumber\\
&&~~~~\times\sum_{\sigmav_{\!\ell^\prime}}\int\!
\mathrm d \Hv_{\ell^\prime}\mathrm d
\hat{\hv}_{\ell^\prime}\int_{-\pi}^{\pi}\mathrm d \hat k_{\ell^\prime} M [\Hv_{\ell^\prime},\hat{\hv}_{\ell^\prime},\sigmav_{\ell^\prime}\vert k_{\ell^\prime},\hat k_{\ell^\prime},\theta]\Big]^{-1}\Bigg\}.
\end{eqnarray}
Now the terms in the sums over $\ell,\ell^\prime$ variables are dependent only on the random connectivity variables $\{k_\ell, k_\ell^\prime\}$, which are independent and distributed according to $P_c(k)$, hence by the law of large numbers we arrive at the result (\ref{eq2:Aintegral}).

\section{Calculation of the Fourier transforms}\label{section2:M}
Here we compute the Fourier transforms $\int\!\mathrm d\hat{\hv}~\rme^{-\rmi\xv.\hat{\hv}}\int_{-\pi}^{\pi}\mathrm d \hat k~\rme^{-\rmi\hat k m}M [\Hv,\hat{\hv},\sigmav\vert k,\hat k,\theta]$ of the function $M$ defined in equation (\ref{def2:M}), where $\xv\in\mathbb{R}^{n}$ and $m\in\mathbb{Z}$. First we expand that part of the exponential function which depends on $Q$, which is defined in equation (\ref{eq2:Q}). Next we integrate out the $\hat k$ variables, which leads us to
\begin{eqnarray}
&&\hspace{-25mm}\int\mathrm d\hat{\hv}~\rme^{-\rmi\xv.\hat{\hv}}\int_{-\pi}^{\pi}\mathrm d \hat k~\rme^{-\rmi\hat k m}M [\Hv,\hat{\hv},\sigmav\vert k,\hat k,\theta]\nonumber\\
&&=\int\mathrm d\hat{\hv}~\rme^{i\hat{\hv}
.\{\Hv-\xv-\thetav\}-\rmi \sum_{\alpha}\hat{D}_{\alpha}(\sigma_\alpha ,H_{\alpha})}\frac{~\rme^{-c}c^{k-m}}{(k-m)!}\tilde Q^{k-m}(\sigmav , \hat{\hv})\label{eq2:comp-of-M}
\end{eqnarray}
In the above we used the short-hand $\tilde Q(\sigmav ,\hat{\hv})=Q(\sigmav ,\hat{\hv},0)+1$. Raising  $\tilde Q$ to the power $k-m$ gives
\begin{eqnarray}
\hspace*{-20mm}
\tilde{Q}^{k-m}(\sigmav ,\hat{\hv})&=&\left[\sum_{\sigmav^\prime}\int\!\mathrm d
\hat{\hv}^\prime \mathrm d \hat{k}^\prime P (\sigmav^\prime\! ,
\hat{\hv}^\prime\!,\hat{k}^\prime)~\rme^{-\rmi\hat{k}^\prime}\Big\langle ~\rme^{-\rmi  J[\hat{\hv} .\sigmav^\prime+\hat{\hv}^{\prime} .\sigmav]}\Big\rangle_{\!J}\right]^{k-m}
\nonumber\\
\hspace*{-20mm}
&=&\prod_{\ell=1}^{k-m}\left[\sum_{\sigmav_\ell}\int\mathrm d
\hat{\hv}_\ell \mathrm d J_\ell P(J_\ell)\int_{-\pi}^{\pi}\mathrm d \hat{k}_\ell P (\sigmav_\ell ,
\hat{\hv}_\ell,\hat{k}_\ell)~\rme^{-\rmi\hat{k}_\ell}~\rme^{-\rmi J_\ell\hat{\hv}_\ell .\sigmav}\right]\nonumber\\
&&\hspace{2mm}\times\rme^{-\rmi  \hat{\hv} .\sum_\ell J_\ell\sigmav_\ell}.\label{eq2:comp-of-Q}
\end{eqnarray}
Now inserting above result into the expression (\ref{eq2:comp-of-M}) and integrating out the $\hat{\hv}$ variables yields equation for the Fourier transform (\ref{eq2:Mft}).
\section{The  joint spin-field probability distributions in equilibrium}\label{section2:EquilibrCalc}
In this section we compute the joint spin-field probability distributions $D$ and $A$ in equilibrium. We note that both can defined via the Fourier transforms (\ref{eq2:Mft}) of the function $M$. First, we consider equation for $D$ (\ref{eq2:D}). Using expression (\ref{eq2:DconjEq}) for the conjugate parameter $\hat{D}_\alpha$ in this equation, combined with the equality (\ref{eq2:FTeq}), gives us
\begin{eqnarray}
\hspace{-15mm}
D(\sigma ,h)&=&\sum_{k\geq0}P_c(k)\frac{1}{M_k}\sum_{\sigmav}\int\!\mathrm d \Hv
\prod_{\ell=1}^{k}\bigg[\sum_{\sigmav_\ell}\int\!\mathrm d
\hat{\hv}_\ell \mathrm d J_\ell P(J_\ell)
\nonumber\\
\hspace*{-15mm}
&&\hspace*{30mm}\times\int_{-\pi}^{\pi}\!\mathrm d \hat{k}_\ell P (\sigmav_\ell ,
\hat{\hv}_\ell,\hat{k}_\ell)\rme^{-\rmi\hat{k}_\ell}\rme^{-\rmi J_\ell\hat{\hv}_\ell .\sigmav}\bigg]
\nonumber\\
\hspace*{-15mm}
&&\times\delta\Big[\Hv-\sum_\ell J_\ell\sigmav_\ell-\thetav\Big]
~\rme^{\frac{1}{2}\beta\sigmav.[ \Hv+\thetav]}
~\delta_{\sigma,\sigma_\gamma}\delta(h-H_{\gamma})\nonumber\\
\hspace*{-15mm}
&=&\sum_{k\geq0}P_c(k)\frac{1}{M_k}\sum_{\sigmav}\int\!\mathrm d \Hv
\prod_{\ell=1}^{k}\left[\sum_{\sigmav_\ell}\int\!\mathrm d J_\ell P(J_\ell)\int_{-\pi}^{\pi}\!\mathrm d \hat{k}_\ell P (\sigmav_\ell ,\hat{k}_\ell)~\rme^{-\rmi\hat{k}_\ell}\right]\nonumber\\
\hspace*{-15mm}
&&\times\delta\Big[\Hv-\sum_\ell J_\ell\sigmav_\ell-\thetav\Big]
~\delta_{\sigma,\sigma_\gamma}\delta(h-H_{\gamma})~\rme^{\beta\sigmav.\Hv}
\end{eqnarray}
where $M_k$ is defined in (\ref{def2:Mk}). Summing and integrating over the variables $\sigma_\gamma$ and $H_\gamma$, respectively, leads us to the equilibrium form (\ref{eq2:Deq}) of the  joint spin-field distribution. In a similar manner we obtain the equilibrium version of $A$, which is given by
\begin{eqnarray}
\hspace*{-20mm}
A[s,s^\prime\!;h,h^\prime\!;\tilde s]&=&
\int\!\mathrm d J~P(J)\frac{1}{Z_A}\sum_{k,k^\prime}\frac{P_c(k)k}{c}\frac{P_c(k^\prime)k^\prime}{c}\frac{1}{M_{k}M_{k^\prime}}
\sum_{\sigmav,\sigmav^\prime}\int\!\mathrm d \Hv\mathrm d \Hv^\prime
\nonumber\\
\hspace*{-20mm}
&&\hspace*{-15mm}
 \times\prod_{\ell=1}^{k-1}\left[\sum_{\sigmav_\ell}\int\!\mathrm d J_\ell P(J_\ell)\int_{-\pi}^{\pi}\!\mathrm d \hat{k}_\ell P (\sigmav_\ell ,\hat{k}_\ell)\rme^{-\rmi\hat{k}_\ell}\right]
~\delta\Big[\Hv\!-\!\sum_\ell J_\ell\sigmav_\ell\!-\!\thetav\!-\!J\sigmav^\prime\Big]
\nonumber\\
\hspace*{-20mm}
&&
\hspace*{-15mm}
 \times\prod_{r=1}^{k^\prime-1}\left[\sum_{\sigmav_r}\int\!\mathrm d J_r P(J_r)\int_{-\pi}^{\pi}\!\mathrm d \hat{k}_r P (\sigmav_r ,
\hat{\hv}_r)\rme^{-\rmi\hat{k}_r}\right]
~\delta\Big[\Hv^\prime\!-\!\sum_r J_r\sigmav_r\!-\!\thetav\!-\!J\sigmav\Big]
\nonumber\\
\hspace*{-20mm}
&&\times\rme^{\beta\sigmav.\Hv+\beta\sigmav^\prime.\Hv^\prime-\beta J\sigmav.\sigmav^\prime}
\nonumber\\
\hspace*{-20mm}
&& \times\delta_{s^\prime,\sigma_1}~\delta_{s,\sigma_1^\prime}~\delta
[h^\prime -H_1]~
\delta [ h -H_1^\prime+2J\tilde
s].\end{eqnarray}
The above result can be written in the form given by equation (\ref{eq2:Aeq}).
\section{Stationary points of the dynamic equation}\label{section2:FixedPoints}
Here we show that the probability distributions $D$ and $A$ in equilibrium are stationary points of our dynamic equation (\ref{eq2:diffusion}). First, we consider that part of (\ref{eq2:diffusion}) which is dependent on the joint spin-field distribution $D(s,h)$ only. Inserting the equilibrium form
(\ref{eq2:Deq}) of this distribution  into the first line in the right-hand side of (\ref{eq2:diffusion}) leads to
\begin{eqnarray}
&&\hspace{-5mm}\frac{1}{2}\left [1+s\tanh[\beta h]\right
]~\rme^{-\beta s h}\Phi[h]-\frac{1}{2}\left [1-s\tanh[\beta h]\right
]~\rme^{\beta s h}\Phi[h]\nonumber\\
&&\hspace{10mm}=\Phi[h]\lbrace -\sinh[\beta s h]+\cosh[\beta s h]\tanh[\beta s h]\rbrace\nonumber\\
&&\hspace{10mm}=0.
\end{eqnarray}
Second, we compute that part of the right-hand side of (\ref{eq2:diffusion}) which is explicitly dependent on the kernel $A[s,s^\prime\!,h,h^\prime\!,\tilde s]$ only. Using our equilibrium form (\ref{eq2:Aeq}) of this kernel  in the last two lines of the right-hand side of (\ref{eq2:diffusion}) results in
\begin{eqnarray}
 &&\hspace{-20mm}\frac{1}{2}c\sum_{s^\prime}\int\!\mathrm{d}h^\prime
[1-s^\prime\tanh[\beta h^\prime]]\Big\langle~\rme^{\beta sh+\beta  s^\prime h^\prime+\beta Jss^\prime }\Lambda[h+Js^\prime ;h^\prime-J s]\Big\rangle_{\!J}
\nonumber\\
&&\hspace{-24mm}-\frac{1}{2}c\sum_{s^\prime}\int\!\mathrm{d}h^\prime
[1-s^\prime \tanh[\beta h^\prime]]\Big\langle~\rme^{\beta sh+\beta s^\prime  h^\prime-\beta Jss^\prime }\Lambda[h-Js^\prime ;h^\prime-J s]\Big\rangle_{\!J}
\nonumber\\
&&\hspace{-10mm}=\frac{1}{2}c ~\rme^{\beta s h}\int\!\mathrm{d} J P(J)\int\!\mathrm{d}h^\prime\Big\{
\nonumber\\
&&\hspace{-7mm}+\left([1\!-\!\tanh[\beta h^\prime]]~\rme^{\beta h^\prime}-[1\!+\!\tanh[\beta h^\prime]]~\rme^{-\beta h^\prime}\right )~\rme^{\beta Js}\Lambda[h+J;h^\prime-J s]\nonumber\\
&&\hspace{-7mm}+\left([1\!+\!\tanh[\beta h^\prime]]~\rme^{-\beta h^\prime} -[1\!-\!\tanh[\beta h^\prime]]~\rme^{\beta h^\prime}\right )~\rme^{-\beta Js}\Lambda[h-J;h^\prime-J s]\Big\}\nonumber\\
&&\hspace{-10mm}=0.
\end{eqnarray}
We conclude that the right-hand side of the dynamic equation (\ref{eq2:diffusion}) is exactly zero for all $s\in\{-1,1\}$ and all $h\in\mathbb{R}$ as soon as the equilibrium relations (\ref{eq2:Deq}) and (\ref{eq2:Aeq}) hold.
\section{RS calculations}\label{section2:RScalc}
In this section we derive an equation for the functional distribution (\ref{def2:Wm}) and compute the replica symmetric versions of the kernels $A$ and $D$. First, we compute the functional distribution $W[\{P\};\vert m]$, where $m\in\mathbb{Z}$. For this we consider the Fourier transform $\int_{-\pi}^{\pi}\!\mathrm d \hat k \rme^{-\rmi \hat k m} P_{RS}(\sigmav, \hat{\hv},\hat k)$ of the RS order parameter function. Using result (\ref{eq2:Mrs}) for $M_{RS}$, and the saddle-point equation (\ref{eq2:P}), we have
\begin{eqnarray}
&&\hspace{-25mm}\int\!\left\lbrace \mathrm d
P\right\rbrace\!\int_{-\pi}^{\pi}\mathrm d \hat k~W[\{P\};\hat k]~\rme^{-\rmi \hat k m}\prod_{\alpha=1}^n
P(\sigma_{\alpha},\hat{h}_{\alpha})
\nonumber\\
&&\hspace{-20mm}=\sum_{k\geq0}P_c(k)\frac{\int\mathrm d \Hv\int_{-\pi}^{\pi}\mathrm d \hat k~\rme^{-\rmi \hat k m}M_{RS} [\Hv,\hat{\hv},\sigmav\vert k,\hat k,\theta]}{\sum_{\sigmav}\int\mathrm d \Hv\mathrm d
\hat{\hv}~M_{RS} [\Hv,\hat{\hv},\sigmav\vert k,\theta]}
\nonumber\\
&&\hspace{-20mm}=\sum_{k\geq0}P_c(k)\frac{k!}{(k\!-\!m)!}c^{-m}\frac{1}{M_k^n}\int\!\prod_{\ell=1}^{k-m}\left[\mathrm d J_\ell P(J_\ell) \left\lbrace \mathrm d
P_\ell\right\rbrace \int_{-\pi}^{\pi}\!\mathrm d \hat{k}_\ell~ W[\{P_\ell\};\hat{k}_\ell]~\rme^{-\rmi \hat{k}_\ell}\right]
\nonumber\\
&&\hspace{-15mm}\times\prod_{\alpha=1}^n\int\!\mathrm d H_\alpha d(\sigma_\alpha,H_\alpha)\rme^{i\hat{h}_\alpha
\{H_\alpha-\theta\}}\prod_{\ell=1}^{k-m}\Big[\sum_{\sigma_{\ell}^{\alpha}}\int\!\mathrm d
\hat{h}_{\ell}^{\alpha}~
P_\ell(\sigma_{\ell}^{\alpha},\hat{h}_{\ell}^{\alpha})\rme^{-\rmi J_\ell[\hat{h}_{\ell}^{\alpha}\sigma_\alpha+\hat{h}_\alpha\sigma_{\ell}^{\alpha}]}\Big]
\nonumber\\
&&\hspace{-20mm}=\int\! \left\lbrace \mathrm d
P\right\rbrace\sum_{k\geq0}P_c(k)\frac{k!}{(k\!-\!m)!}c^{-m}\int\!\prod_{\ell=1}^{k-m}\left[\mathrm d J_\ell P(J_\ell) \left\lbrace \mathrm d
P_\ell\right\rbrace\! \int_{-\pi}^{\pi}\!\mathrm d \hat{k}_\ell~ W[\{P_\ell\};\hat{k}_\ell]\rme^{-\rmi \hat{k}_\ell}\right]
\nonumber\\
&&\hspace{-18mm}\times\prod_{\sigma,\hat h}\delta\!\left[\!P(\sigma,\hat h)\!-\!\frac{\int\!\mathrm d H d(\sigma,H)\rme^{i\hat{h}
\{H-\theta\}}\!\prod_{\ell=1}^{k-m}\!\left[\!\sum_{\sigma_{\ell}}\!\int\!\mathrm d
\hat{h}_{\ell}
P_\ell(\sigma_{\ell},\hat{h}_{\ell})\rme^{-\rmi J_\ell[\hat{h}_{\ell}\sigma+\hat{h}\sigma_{\ell}]}\right]}{Z[\{P_1,\ldots,P_{k-m}\}]}\right]\nonumber\\
&&\hspace{-15mm}\times \frac{1}{M_k^n}Z[\{P_1,\ldots,P_{k-m}\}]^n
~\prod_{\alpha=1}^n
P(\sigma_{\alpha},\hat{h}_{\alpha})\label{eq2:comp-of-W}
\end{eqnarray}
where we have used the short-hands
\begin{eqnarray}
&&\hspace{-25mm}M_k^n=\int\!\prod_{\ell=1}^{k}\left[\mathrm d J_\ell P(J_\ell) \left\lbrace \mathrm d
P_\ell\right\rbrace\!\int_{-\pi}^{\pi}\mathrm d \hat{k}_\ell W[\{P_\ell\};\hat{k}_\ell]~\rme^{-\rmi \hat{k}_\ell}\right]
\nonumber\\
&&\hspace{-16mm}\times\left[\sum_{\sigma}\int\!\mathrm d H \mathrm d\hat{h} ~d(\sigma,H)\rme^{\rmi\hat{h}
(H-\theta)}\prod_{\ell=1}^{k}\Big(\sum_{\sigma_{\ell}}\int\!\mathrm d
\hat{h}_{\ell}
P_\ell(\sigma_{\ell},\hat{h}_{\ell})\rme^{-\rmi J_\ell[\hat{h}_{\ell}\sigma+\hat{h}\sigma_{\ell}]}\Big)\right]^n\label{def2:Mkn}~~
\end{eqnarray}
and
\begin{eqnarray}
\hspace{-20mm}Z[\{P_1,\ldots,P_{k-m}\}]&=&
\sum_{\sigma}\int\!\mathrm d H \mathrm d\hat{h} d(\sigma,H)\rme^{\rmi\hat{h}(H-\theta)}
\nonumber\\
\hspace*{-20mm}
&&\times \prod_{\ell=1}^{k-m}\Big[\sum_{\sigma_{\ell}}\!\int\!\mathrm d
\hat{h}_{\ell}
P_\ell(\sigma_{\ell},\hat{h}_{\ell})\rme^{-\rmi J_\ell[\hat{h}_{\ell}\sigma+\hat{h}\sigma_{\ell}]}\Big]
\nonumber\\
\hspace*{-20mm}
&&\hspace*{-15mm}=~ 2\pi\sum_{\sigma}\!\prod_{\ell=1}^{k-m}\Big[
\sum_{\sigma_{\ell}}\!\int\!\mathrm d \hat{h}_{\ell}
P_\ell(\sigma_{\ell},\hat{h}_{\ell})\rme^{- \rmi J_\ell\hat{h}_{\ell}
\sigma}\Big] d\big(\sigma,\!\sum_{\ell}
J_\ell\sigma_{\ell}\!+\!\theta\big)\label{def2:Z}.
\end{eqnarray}
Solving equation (\ref{eq2:comp-of-W}) for the functional distribution $\int_{-\pi}^{\pi}\!\mathrm d \hat k~W[\{P\};\hat k]~\rme^{-\rmi \hat k m}$, followed by taking the replica limit $n\rightarrow0$ in the functions $M_k^n$ and $Z^n$ of the resulting expression, then leads to equation (\ref{eq2:W}).

Second, we compute the RS joint spin-field probability distribution $D(s,h)$.  Using the saddle-point equation (\ref{eq2:D}) for this distribution, combined with the result (\ref{eq2:Mrs}) for $M_{RS}$, applied to $m=0$, gives us
\begin{eqnarray}
\hspace{-15mm}D(\sigma ,h)&=&\sum_{k\geq0}P_c(k)\frac{\sum_{\sigmav}\int\!\mathrm d \Hv\mathrm d
\hat{\hv}~M_{RS} [\Hv,\hat{\hv},\sigmav\vert k,\theta]~\delta_{\sigma,\sigma_\gamma}\delta(h-H_{\gamma})}{\sum_{\sigmav}\int\!\mathrm d \Hv\mathrm d
\hat{\hv}~M_{RS} [\Hv,\hat{\hv},\sigmav\vert k,\theta]}
\nonumber\\
\hspace*{-15mm}&=& \sum_{k\geq0}P_c(k)\frac{1}{M_k^n}~\int\!\prod_{\ell=1}^{k}\left[\mathrm d J_\ell P(J_\ell) \left\lbrace \mathrm d
P_\ell\right\rbrace~ \int_{-\pi}^{\pi}\!\mathrm d \hat{k}_\ell W[\{P_\ell\};\hat{k}_\ell]~\rme^{-\rmi \hat{k}_\ell}\right]
\nonumber\\
\hspace*{-15mm}
&&\times\frac{\!\prod_{\ell=1}^{k}\left[
\sum_{\sigma_{\ell}}\int\!\mathrm d \hat{h}_{\ell}
P_\ell(\sigma_{\ell},\hat{h}_{\ell})\rme^{- i J_\ell\hat{h}_{\ell}
\sigma}\right] d(\sigma,h)~\delta(h-\!\sum_{\ell}
J_\ell\sigma_{\ell}\!-\!\theta)}{\sum_{\sigma}\!\prod_{\ell=1}^{k}\left[
\sum_{\sigma_{\ell}}\int\!\mathrm d \hat{h}_{\ell}
P_\ell(\sigma_{\ell},\hat{h}_{\ell})\rme^{- i J_\ell\hat{h}_{\ell}
\sigma}\right] d\big(\sigma,\!\sum_{\ell}
J_\ell\sigma_{\ell}\!+\!\theta\big)}\nonumber\\
\hspace*{-15mm}
&&\times Z[\{P_1,\ldots,P_{k}\}]^{n}\label{eq2:comp-of-D}
\end{eqnarray}
where the functions $M_k^n$ and $Z[\ldots]^{n}$ are defined by (\ref{def2:Mkn}) and (\ref{def2:Z}) respectively. Taking the replica limit in equation (\ref{eq2:comp-of-D}) leads us to the result (\ref{eq2:Drs}).

Finally, we compute the RS version of the kernel (\ref{eq2:A}). We consider numerator and denominator in the average over the vertex connectivities in this equation separately. Using equality (\ref{eq2:Mrs}) for the function $M_{RS}$  we obtain the numerator
\begin{eqnarray}
\hspace{-20mm}{\rm num}&=&
\sum_{\sigmav,\sigmav^\prime}\int\!\mathrm d \Hv\mathrm d \Hv^\prime\mathrm d
\hat{\hv}\mathrm d
\hat{\hv}^\prime M_{RS} [\Hv,\hat{\hv},\sigmav\vert k-1,\theta]~M_{RS} [\Hv^\prime,\hat{\hv}^\prime,\sigmav^\prime\vert k^\prime-1,\theta]
\nonumber\\
\hspace{-20mm}&&
\times\delta_{s^\prime ,\sigma_1}\delta_{s,\sigma_1^\prime}\delta
[h^\prime  -H_1]
 \left\langle\delta [ h -H_1^\prime+2J\tilde
s]e^{-\rmi  J[\hat{\hv} .\sigmav^\prime+
\hat{\hv}^{\prime} .\sigmav]}\right\rangle_J
\nonumber\\
\hspace{-20mm}&=&
\frac{e^{-c}c^{k-1}}{(k-1)!}\int\!\prod_{\ell=1}^{k-1}\left[\mathrm d J_\ell P(J_\ell) \left\lbrace \mathrm d
P_\ell\right\rbrace~ \int_{-\pi}^{\pi}\mathrm d \hat{k}_\ell W[\{P_\ell\};\hat{k}_\ell]~\rme^{-\rmi \hat{k}_\ell}\right]
\nonumber\\
\hspace{-20mm}
&&\times\frac{e^{-c}c^{k^\prime-1}}{(k^\prime-1)!}\int\!\prod_{r=1}^{k^\prime-1}\left[\mathrm d J^{\prime}_r P(J^{\prime}_r) \left\lbrace \mathrm d
Q_r\right\rbrace~ \int_{-\pi}^{\pi}\mathrm d \hat{k}_r W[\{Q_r\};\hat{k}_r]~\rme^{-\rmi \hat{k}_r}\right]
\nonumber\\
\hspace{-20mm}
&&
\times\Bigg\langle\sum_{\sigma,\sigma^\prime}\prod_{\ell=1}^{k-1}\left[\sum_{\sigma_{\ell}}\int\mathrm d
\hat{h}_{\ell}
P_\ell(\sigma_{\ell},\hat{h}_{\ell})\rme^{-\rmi J_\ell\hat{h}_{\ell}\sigma}\right]~d(\sigma,\sum_\ell J_\ell\sigma_\ell+\theta+J\sigma^\prime)
\nonumber\\
\hspace{-20mm}&&
\times\prod_{r=1}^{k^\prime-1}\left[\sum_{\sigma_{r}}\int\mathrm d
\hat{h}_{r}
Q_r(\sigma_{r},\hat{h}_{r})\rme^{-\rmi J^{\prime}_r\hat{h}_{r}\sigma^{\prime}}\right]~d(\sigma^{\prime},\sum_r J^{\prime}_r\sigma_r+\theta+J\sigma)
\nonumber\\
\hspace{-20mm}&&
\times\delta_{s^\prime ,\sigma}\delta_{s,\sigma^\prime}~\delta
[h^\prime -\sum_\ell J_\ell\sigma_\ell-\theta-J\sigma^\prime]~
 \delta [ h -\sum_r J^{\prime}_r\sigma_r-\theta-J\sigma+2J\tilde
s]\nonumber\\
\hspace{-20mm}&&
\times\Bigg\{\sum_{\sigma,\sigma^\prime}\prod_{\ell=1}^{k-1}\left[\sum_{\sigma_{\ell}}\int\mathrm d
\hat{h}_{\ell}
P_\ell(\sigma_{\ell},\hat{h}_{\ell})\rme^{-\rmi J_\ell\hat{h}_{\ell}\sigma}\right]d(\sigma,\sum_\ell J_\ell\sigma_\ell+\theta+J\sigma^\prime)
\nonumber\\
\hspace{-20mm}&&
\times\!\prod_{r=1}^{k^\prime-1}\!\left[\sum_{\sigma_{r}}\int\mathrm d
\hat{h}_{r}
Q_r(\sigma_{r},\hat{h}_{r})\rme^{-\rmi J^{\prime}_r\hat{h}_{r}\sigma^{\prime}}\right]d(\sigma^{\prime},\!\sum_r J^{\prime}_r\sigma_r+\theta+J\sigma)\Bigg\}^{n-1}\Bigg\rangle_J
\end{eqnarray}
and the denominator
\begin{eqnarray}
\hspace{-20mm}
{\rm den}&=&
\sum_{\sigmav,\sigmav^\prime}\int\!\mathrm d \Hv\mathrm d \Hv^\prime\mathrm d
\hat{\hv}\mathrm d
\hat{\hv}^\prime M_{RS} [\Hv,\hat{\hv},\sigmav\vert k,\theta]~M_{RS} [\Hv^\prime,\hat{\hv}^\prime,\sigmav^\prime\vert k^\prime,\theta]
\nonumber\\
\hspace{-20mm}
&=&
\frac{e^{-c}c^{k}}{k!}\int\!\prod_{\ell=1}^{k}\left[\mathrm d J_\ell P(J_\ell) \left\lbrace \mathrm d
P_\ell\right\rbrace~ \int_{-\pi}^{\pi}\mathrm d \hat{k}_\ell W[\{P_\ell\};\hat{k}_\ell]~\rme^{-\rmi \hat{k}_\ell}\right]
\nonumber\\
\hspace{-20mm}&&\times\left[\sum_{\sigma}\prod_{\ell=1}^{k}\left[\sum_{\sigma_{\ell}}\int\mathrm d
\hat{h}_{\ell}
P_\ell(\sigma_{\ell},\hat{h}_{\ell})\rme^{-\rmi J_\ell\hat{h}_{\ell}\sigma}\right]~d(\sigma,\sum_\ell J_\ell\sigma_\ell+\theta)\right]^n
\nonumber\\
\hspace{-20mm}
&&\times\frac{e^{-c}c^{k^\prime}}{k^\prime!}\int\!\prod_{r=1}^{k^\prime}\left[\mathrm d J^{\prime}_r P(J^{\prime}_r) \left\lbrace \mathrm d
Q_r\right\rbrace~ \int_{-\pi}^{\pi}\mathrm d \hat{k}_r W[\{Q_r\};\hat{k}_r]~\rme^{-\rmi \hat{k}_r}\right]
\nonumber\\
\hspace{-20mm}&&
\times\left[\sum_{\sigma^\prime}\prod_{r=1}^{k^\prime}\left[\sum_{\sigma_{r}}\int\mathrm d
\hat{h}_{r}
Q_r(\sigma_{r},\hat{h}_{r})\rme^{-\rmi J^{\prime}_r\hat{h}_{r}\sigma^{\prime}}\right]~d(\sigma^{\prime},\sum_r J^{\prime}_r\sigma_r+\theta)\right]^n.
\end{eqnarray}
Combining these latter two results in (\ref{eq2:A}) and taking the $n\rightarrow0$ replica limit gives equation (\ref{eq2:Ars}).

\section{Dynamic equation for the Ising ferromagnet with dilution}\label{section2:ODE}
Here we show that the macroscopic equation (\ref{eq2:diffusion}) for the Ising spin system governed by (\ref{def2:Ediluted}) can be reduced to a system of ordinary differential equations. In the present Ising ferromagnet with dilution (\ref{def2:Ediluted}) the fields (\ref{eq2:field}) can take only discrete values, which implies that the distributions (\ref{def2:jsfield}) and (\ref{def2:A}) can be written in the form (\ref{def2:Psn}) and (\ref{def2:Asn}) respectively. Inserting (\ref{def2:Psn}) and (\ref{def2:Asn}) into both sides of (\ref{eq2:diffusion}) gives
\begin{eqnarray}
  &&\hspace{-20mm}
  \frac{\rmd}{\rmd t}\sum_{n=-k}^{k}P_t(s,n)~\delta (h\!-\!Jn\!-\!\theta)
  ~=~ \frac{1}{2}\left [1+s\tanh[\beta h]\right
]\sum_{n=-k}^{k}P_t(-s,n)~\delta (h\!-\!Jn\!-\!\theta)
\nonumber\\
&&\hspace*{29mm}
-\frac{1}{2}\left [1-s\tanh[\beta h]\right
]\sum_{n=-k}^{k}P_t(s,n)~\delta (h\!-\!Jn\!-\!\theta)
\nonumber\\
&&
\hspace{-9mm}+\frac{1}{2}k\sum_{s^{\prime}}\int\!\mathrm{d} h^{\prime}
[1-s^{\prime}\tanh[\beta h^{\prime}]]\sum_{n=-k+1}^{k-1}\sum_{n^\prime=-k+1}^{k-1}
\nonumber\\
&&\hspace{1mm}\times \left\langle A_t[s,s^{\prime}\!;n,n^{\prime}\vert\tau]~\delta [h^{\prime}\! -\!Jn^{\prime}\!-\!\theta\!-\!J\tau s]~
\delta [ h\! -\!J n\!-\!\theta\!+\!J\tau s^{\prime}]\right\rangle_\tau\nonumber\\[1mm]
&&\hspace{-9mm}-\frac{1}{2}k\sum_{s^{\prime}}\int\!\mathrm{d}h^{\prime}
[1-s^{\prime}\tanh[\beta h^{\prime}]]\sum_{n=-k+1}^{k-1}\sum_{n^\prime=-k+1}^{k-1}
\nonumber\\
&&\hspace{1mm}\times \left\langle A_t[s,s^{\prime}\!;n,n^{\prime}\vert\tau]~
\delta [h^{\prime}\! -\!Jn^{\prime}\!-\!\theta\!-\!J\tau s]~\delta [ h\! -\!J n\!-\!\theta\!-\!J\tau s^{\prime}]\right\rangle_\tau\label{eq2:comp-of-ODE1}
\end{eqnarray}
in which the averages over $\tau$ refer to the distribution $P(\tau)=p\delta_{\tau,1}+(1-p)\delta_{\tau,0}$.
We move the time derivative inside the sum on the left of the above equation. On the right side we average over $\tau$,
take the sums over $s^\prime$, and integrate out $h^\prime$ variables.
These manipulations produce
 \begin{eqnarray}
 &&\hspace{-25mm}\sum_{n=-k}^{k}\frac{\mathrm{d}}{\mathrm{d} t}P_t(s,n)~\delta (h\!-\!Jn\!-\!\theta)~=~ \frac{1}{2}\left [1\!+\!s\tanh[\beta (Jn\!+\!\theta)]\right
]\sum_{n=-k}^{k}P_t(-s,n)~\delta (h\!-\!Jn\!-\!\theta)
\nonumber\\
&&\hspace{22mm}-\frac{1}{2}\left [1\!-\!s\tanh[\beta (Jn\!+\!\theta)]\right
]\sum_{n=-k}^{k}P_t(s,n)~\delta (h\!-\!Jn\!-\!\theta)
\nonumber\\%
&&\hspace*{-15mm}
+\sum_{n=-k}^{k-2}\sum_{n^\prime=-k+1}^{k-1}\frac{1}{2}kp~[1\!-\!\tanh[\beta J(n^{\prime}\!+\!s)\!+\!\beta\theta]]
A_t[s,1;n\!+\!1,n^{\prime}\vert1]\delta ( h\! -\!J n\!-\!\theta)
\nonumber\\
&&\hspace{-15mm}
+\!\!\sum_{n=-k+2}^{k}\sum_{n^\prime=-k+1}^{k-1}\frac{1}{2}kp~[1\!+\!\tanh[\beta J(n^{\prime}\!+\!s)\!+\!\beta\theta]]
A_t[s,\!-\!1;n\!-\!1,n^{\prime}\vert1]\delta ( h\! -\!J n\!-\!\theta)
\nonumber\\
&&\hspace{-15mm}
-\sum_{n=-k}^{k-2}\sum_{n^\prime=-k+1}^{k-1}\frac{1}{2}kp~[1\!+\!\tanh[\beta J(n^{\prime}\!+\!s)\!+\!\beta\theta]]
A_t[s,\!-\!1;n\!+\!1,n^{\prime}\vert1]\delta ( h\! -\!J n\!-\!\theta)
\nonumber\\
&&\hspace{-15mm}
-\!\!\sum_{n=-k+2}^{k}\sum_{n^\prime=-k+1}^{k-1}\frac{1}{2}kp~[1\!-\!\tanh[\beta J(n^{\prime}\!+\!s)\!+\!\beta\theta]]
A_t[s,1;n\!-\!1,n^{\prime}\vert1]\delta ( h\! -\!J n\!-\!\theta).
\nonumber\\&&\label{eq2:comp-of-ODE2}
\end{eqnarray}
The result (\ref{eq2:ODE}) follows immediately from the above equation.
\section{Initial conditions}\label{section2:initial}
In this appendix we compute the relevant initial conditions for the system of ordinary equations (\ref{eq2:ODE}). We choose an initial microscopic state of the system in which each spin $\sigma_i$ is drawn randomly and independently according to $P_0(\sigma_i)=\frac{1}{2}(1+\sigma_i m_0) $, where $m_0\in[-1,1]$ is the prescribed initial magnetization of the whole system, i.e.
\begin{eqnarray}
P_0(\sigmav) &=& \prod_{i=1}^N\frac{1}{2}(1+ \sigma_i m_0)\label{eq2:Prob0}.
\end{eqnarray}
Given (\ref{eq2:Prob0}), the spin-field probability distribution $P_0(s,n)$ for large Ising ferromagnets defined on random graphs with vertex degree distribution $P_c(k^\prime)$ is given by
\begin{eqnarray}
P_0(s,n)&=& \lim_{N\rightarrow\infty}\sum_{\sigmav}P_0(\sigmav)\frac{1}{N}\sum_i^N\delta_{s,\sigma_i}\delta_{n, \sum_{j\neq i}c_{ij}\sigma_j}
\nonumber \\
&=& \frac{1}{2}(1\!+\! s m_0)\!\sum_{k^\prime\geq 0}\!P_c(k^\prime)\!
\prod_{\ell=1}^{k^\prime}\!\left[\sum_{\sigma_{\ell}}\frac{1}{2}(1\!+\! \sigma_\ell m_0)\right]\!\delta_{n,\sum_{\ell=1}^{k^\prime} \sigma_{\ell}}.
\label{eq2:Psn0}
\end{eqnarray}
For the model (\ref{def2:Ediluted}) in particular, where $k$ is the connectivity of the random regular graph and $p$ is the dilution, the vertex degree distribution is binomial
\begin{eqnarray}
P_c(k^\prime)&=&\Big(\!\begin{array}{c}k\\k^\prime\end{array}
\!\Big)p^{k^\prime}(1-p)^{k-k^\prime}.
\end{eqnarray}
Solving equations (\ref{eq2:Psn0}) and (\ref{eq2:Psn}) for the functional distribution $\tilde W[\{\hat P\}\vert1]$ and the function $d(s,Jn+ \theta)$ then gives
 \begin{eqnarray}
 \tilde W[\{\hat P\}\vert1]&=&\prod_{\sigma,\sigma^\prime}\delta\Big[\hat P(\sigma\vert \sigma^\prime)- \frac{1}{2}(1+ \sigma m_0)\Big]\label{eq2:W0}\\
d(s,Jn+ \theta)&=&\frac{1}{2}(1+ s m_0)\label{eq2:d0}
 \end{eqnarray}
which is the trivial solution of equation (\ref{eq2:WdilFtbin}).

\section{Population dynamics}\label{section2:pop}
The joint spin-field probability distribution $P_t(s,n)$ of the diluted ferromagnet (\ref{def2:Ediluted}) evolves in time according to the system of ordinary differential equations (\ref{eq2:ODE}). Solving this system requires computation of the kernel (\ref{eq2:Asn}), which is dependent on the functional distribution $\tilde W$ and the function $d$ (the order parameters). The saddle-point equations (\ref{eq2:WdilFt}) and (\ref{eq2:Psn}) establish relations between these parameters and their dependence on $P_t(s,n)$. However, solving these equations analytically is generally ruled out,  and one has to solve them numerically using population dynamics \cite{BetheSG}.

 The population dynamics algorithm was also used in the preceding version of the dynamical replica theory, as developed for Poissonian random graphs \cite{MozeikaCoolen}. Here, however, we take an approach which is slightly different from the one in \cite{MozeikaCoolen}. We note that in our dynamical theory we use $P_t(s,n)$  to estimate the order parameters $\tilde W$ and $d$. In particular, the values of the order parameters are considered to be 'good' when the saddle-point equation (\ref{eq2:WdilFt}) for the functional distribution $\tilde W$ is satisfied, and the probability distribution $P(s,n)$ which is computed via saddle-point equation (\ref{eq2:Psn}) equals the instantaneous distribution $P_t(s,n)$. This suggests that the change made by any numerical algorithm to the order parameters $\tilde W$ and $d$ has to reduce the `distance' between the distributions $P_t(s,n)$ and $P(s,n)$, subject to the constraints (\ref{eq2:WdilFt}) and (\ref{eq2:Psn}). The Kullback-Leibler (KL) divergence
\begin{eqnarray}
D_{KL}(P_t\vert\vert P)=\sum_s\sum_n P_t(s,n)\log\left[ \frac{P_t(s,n)}{P(s,n)}\right]\label{def2:KLdist}
\end{eqnarray}
 can play the role of a distance in this context, and we may use e.g. a gradient descent algorithm
 to minimize this distance, viz.
\begin{eqnarray}
\frac{\mathrm d}{\mathrm{d} \epsilon}d(s,Jn+\theta)=-\frac{\partial}{\partial d(s,Jn+\theta)}D_{KL}(P_t\vert\vert P)\label{def2:gradient}
\end{eqnarray}
where $\epsilon$ defines an `algorithmic time'.
To solve equations (\ref{eq2:WdilFt}) and (\ref{eq2:Psn}), we use a combination of both population dynamics and gradient descent. To implement the population dynamics we create a population of $\mathcal{N}$ $2\!\times\!2$ matrices  $\hat P_i(\sigma \vert \sigma^\prime)$, where $i=1\ldots \mathcal{N}$, and we initialize the function
$d(s,Jn+ \theta)$ for $s\in\{-1,1\}$ and $n\in\{-k,\ldots,k\}$. The initial values of population $\lbrace\hat P_i(\sigma \vert \sigma^\prime)\rbrace$ and function $d(s,Jn+ \theta)$ are set to (\ref{eq2:W0}) and (\ref{eq2:d0}), respectively, at $t=0$. For $t>0$ we simply reuse values from the previous time step. We then update the population of matrices and the numbers $d(s,Jn+ \theta)$ until they are stationary, via the following process:
\begin{enumerate}
  \item a number $k^\prime$ is drawn from the binomial distribution $B_{k-1}(k^\prime)$ (\ref{def2:binomial})
  \item $k^\prime$ members $\hat{P}_i(\sigma \vert \sigma^\prime)$ are selected randomly and independently from the population
  \item a new value for $P(\sigma \vert \sigma^\prime)$ is calculated according to
\begin{eqnarray}
 &&\hspace{-15mm}\hat P_{\rm new}(\sigma \vert \sigma^\prime)=\frac{\prod_{l=1}^{k^\prime}\left\lbrace \sum_{\sigma_{l}}
\hat P_l(\sigma_{l}\vert\sigma)\right\rbrace d(\sigma,J\sum_{l=1}^{k^\prime} \sigma_{l}+\theta+J\sigma^\prime)}{\sum_{\sigma^{\prime\prime}}\prod_{l=1}^{k^\prime}\left\lbrace \sum_{\sigma_{l}}
\hat P_l(\sigma_{l}\vert\sigma^{\prime\prime})\right\rbrace d(\sigma^{\prime\prime},J\sum_{l=1}^{k^\prime} \sigma_{l}+\theta)}
\label{eq2:Pupdate}
\end{eqnarray}
  \item a member of the population is selected randomly, and replaced with the newly computed value $\hat P_{\rm new}(\sigma \vert \sigma^\prime)$
  \item a new function $d(s,n)$ is computed according to
\begin{eqnarray}
&&\hspace{-15mm}d_{\rm new}(s,n)=d(s,n)+\Delta\epsilon\frac{d(s,n)}{1+d(s,n)^2}\left[P_t(s,n)-P(s,n)\right]\label{eq2:dUpdate}
\end{eqnarray}
where $0<\Delta\epsilon\ll1$, and $P(s,n)$ is computed according to (\ref{eq2:Psn}) by averaging over the instantaneous values of the population.
\end{enumerate}
The rule (\ref{eq2:dUpdate}) used to update $d(s,n)$ can be regarded as an approximation of the gradient descent equation (\ref{def2:gradient}), which can be derived as follows. First we use the definition of the KL divergence (\ref{def2:KLdist}) and equation (\ref{eq2:Psn}) for $P(s,n)$ to compute the partial derivative in (\ref{def2:gradient}), giving (with the short-hand $d(s,Jn+\theta)\to d(s,n)$)
\begin{eqnarray}
&&
\hspace{-25mm}\frac{\partial}{\partial d(s,n)}D_{KL}(P_t\vert\vert P)
=-\frac{P_t(s,n)}{d(s,n)}
\nonumber
\\
&&\hspace*{-9mm}
+\sum_{s^{\prime}}\sum_{n^{\prime}} P_t(s^\prime,n^\prime)\frac{d(s^\prime,n^\prime)}{P(s^\prime,n^\prime)}\sum_{\tau_1,\ldots,\tau_{k}}P(\tau_1,\ldots,\tau_{k})\int\prod_{\ell=1}^{k}\left[ \left\lbrace \mathrm d \hat
P_\ell\right\rbrace\tilde W[\{\hat P_\ell\}\vert 1]\right]\nonumber\\
&&\hspace{2mm}\times\prod_{\ell=1}^{k}\Big[
\sum_{\sigma_{\ell}^\prime}
\hat P_\ell(\sigma_{\ell}^\prime\vert \tau_\ell s^\prime)\Big] \delta_{n^\prime,\sum_{\ell=1}^k
\tau_\ell\sigma_{\ell}^\prime}~.\prod_{\ell=1}^{k}\Big[
\sum_{\sigma_{\ell}}
\hat P_\ell(\sigma_{\ell}\vert \tau_\ell s)\Big] \delta_{n,\sum_{\ell=1}^k
\tau_\ell\sigma_{\ell}}\nonumber\\
&&\hspace{2mm}\times\left[\sum_{\sigma}\prod_{\ell=1}^{k}\Big[
\sum_{\sigma_{\ell}}\hat P_\ell(\sigma_{\ell}\vert \tau_\ell\sigma)\Big] d\big(\sigma,\!J\sum_{\ell}
\tau_\ell\sigma_{\ell}\!+\!\theta\big)\right]^{-2}.
\label{eq2:comp-of-derivative}
\end{eqnarray}
The result (\ref{eq2:comp-of-derivative}) takes a very simple form when there is no disorder and the distribution $W[\{\hat P\}\vert 1]$ is a functional delta, where one is led to
\begin{eqnarray}
\frac{\mathrm d}{\mathrm{d} \epsilon}d(s,n)&=&\frac{1}{d(s,n)}\left[P_t(s,n)-P(s,n)\right].\label{eq2:approx}
\end{eqnarray}
 To reduce computational costs we use in our population dynamics algorithm approximation (\ref{eq2:approx}), rather than the full version of the gradient descent (\ref{def2:gradient}) which would have required computation of (\ref{eq2:comp-of-derivative}). First, however, expression (\ref{eq2:approx}) is slightly modified according to $1/d(s,n)\rightarrow d(s,n)/[1+d(s,n)^2]$, to prevent unbounded increasing (or decreasing) of $\Delta \epsilon$ in the discrete version of (\ref{eq2:approx}). The number of iterations required to solve saddle-point equations (\ref{eq2:WdilFt},\ref{eq2:Psn}) by the algorithm presented in this section was found to be typically of order $10^2\mathcal{N}$,  for the population size $\mathcal{N}=10^4$.

\section*{References}


\end{document}